\documentclass[]{aa}

\usepackage[varg]{txfonts}
\usepackage{graphicx}

\usepackage{natbib,twoopt}
\usepackage[breaklinks=true]{hyperref} 
\bibpunct{(}{)}{;}{a}{}{,} 
\makeatletter
\newcommandtwoopt{\citeads}[3][][]{\href{http://adsabs.harvard.edu/abs/#3}%
{\def\hyper@linkstart##1##2{}%
\let\hyper@linkend\@empty\citealp[#1][#2]{#3}}}
\newcommandtwoopt{\citepads}[3][][]{\href{http://adsabs.harvard.edu/abs/#3}%
{\def\hyper@linkstart##1##2{}%
\let\hyper@linkend\@empty\citep[#1][#2]{#3}}}
\newcommandtwoopt{\citetads}[3][][]{\href{http://adsabs.harvard.edu/abs/#3}%
{\def\hyper@linkstart##1##2{}%
\let\hyper@linkend\@empty\citet[#1][#2]{#3}}}
\newcommandtwoopt{\citeyearads}[3][][]%
{\href{http://adsabs.harvard.edu/abs/#3}
{\def\hyper@linkstart##1##2{}%
\let\hyper@linkend\@empty\citeyear[#1][#2]{#3}}}
\makeatother

\begin{document}

\title{Investigating the rotational evolution of young, low mass
stars using Monte Carlo simulations}

\author{M. J. Vasconcelos\inst{\ref{UESC},\ref{IPAG}} \and 
J. Bouvier\inst{\ref{IPAG}}}

\institute{LATO - DCET, Universidade Estadual de Santa Cruz, UESC, \\
Rodovia Jorge Amado, km 16, Ilh\'eus/BA, 45662-900, Brazil
\email{mjvasc@uesc.br}\label{UESC}
\and
Univ. Grenoble Alpes, IPAG, F-38000 Grenoble, France \\
CNRS, IPAG, F-38000 Grenoble, France\label{IPAG}}

\date{Received / Accepted}

\abstract {Young stars rotate well below break-up velocity, which
is thought to result from the magnetic coupling with their accretion
disk.} {We investigate the rotational evolution of young stars under
the disk locking hypothesis through Monte Carlo simulations.} {Our
simulations include 280,000 stars, each of which is initially
assigned a mass, a rotational period, and a mass accretion rate.
The mass accretion rate depends on both mass and time, following
power-laws indices 1.4 and -1.5, respectively. A mass-dependent
accretion threshold is defined below which a star is considered as
diskless, which results in a distribution of disk lifetimes that
matches observations. Stars are evolved at constant angular spin
rate while accreting and at constant angular momentum when they
become diskless.} {Starting with a bimodal distribution of periods
for disk and diskless stars, we recover the bimodal period distribution
seen in several young clusters. The short period peak consists
mostly of diskless stars and the long period one is mainly populated
by accreting stars. Both distributions, however, present a long
tail towards long periods and a population of slowly rotating {\it
diskless} stars is observed at all ages. We reproduce the observed
correlations between disk fraction and spin rate, as well as between
IR excess and rotational period. The period-mass relation we derive
from the simulations exhibits the same global trend as observed in
young clusters only if we release the disk locking assumption for
the lowest mass stars.  Finally, we find that the time evolution
of median specific angular momentum follows a power law index of
-0.65 for accreting stars, as expected from disk locking, and of
-0.53 for diskless stars,  a shallower slope that results from a
wide distribution of disk lifetimes. At the end of the accretion
phase, our simulations reproduce the angular momentum distribution
of the low-mass members of the 13 Myr h Per cluster.}{ Using
observationally-documented distributions of disk lifetimes, mass
accretion rates, and initial rotation periods, and evolving an
initial population from 1 to 12 Myr, we reproduce the main
characteristics of pre-main sequence angular momentum evolution,
which supports the disk locking hypothesis.} {}


\titlerunning{Investigating the rotational evolution of young, low mass
stars} \authorrunning{Vasconcelos \& Bouvier}

\maketitle

\section{Introduction}

Several determinations of the rotational properties of young clusters
have shown that the stellar period changes as a function of time
\citepads[e.g.,][for a review]{2013EAS....62..143B}. The distribution
of rotational periods of the Orion Nebula Cluster with an age of 2
Myr presents values that range from less than 1 day to $> 15$ days.
The older cluster NGC 2547, which is 40 Myr old, shows almost the
same amplitude of periods but with lower mass stars (M$_\ast \lesssim
0.5$ M$_\odot$) rotating faster in average. Main sequence (MS)
stars, like the Sun, rotate much more slowly than young stars.

According to several studies (e.g., \citeads{2006ApJ...646..297R};
\citeads{2007ApJ...671..605C}), during the first few Myr of evolution,
stars showing evidence of the presence of disks like infrared excess
emission, broad H$\alpha$ lines, excess UV emission, etc., rotate
slower in average than diskless stars. These observations are
explained by the disk locking hypothesis: during the time the star
remains attached to the disk, the stellar rotation rate remains
constant.  As soon as the star loses its disk, it starts to spin-up,
due to the expected radius contraction on the Hayashi track. According
to \citetads{2013A&A...556A..36G}, from the Zero Age Main Sequence
(ZAMS) to the MS, the star does not conserve its angular momentum
which can be lost via magnetic stellar winds, and eventually all
stars end up on the MS with the same angular velocity.

There are, however, some open issues related to this picture. Based
on the disk frequency in several young clusters,
\citetads{2009AIPC.1158....3M} pointed out that the disk fraction
as a function of time falls off exponentially. The mean disk lifetime
is expected to be around 2 - 3 Myr, although some results points
to longer values \citepads[5 - 6 Myr,][]{2013MNRAS.434..806B}.
Measurements of the mass accretion rate in clusters of different
ages point to a dependency with the age $t$ of the star and its
mass $M_\ast$ in the form $\dot{M}_\mathrm{acc} \propto t^{-\eta}
M_\ast^\mathrm{b}$. However, there is no general agreement about
the value of the exponents. According to the self-similar accretion
theory \citepads{1998ApJ...495..385H}, the mass accretion rate is
independent of the mass and $\eta = 3/2$ or at least $\eta > 1$.
Several works provide different values for $\eta$ and b. For example,
\citetads{2012ApJ...755..154M} analyzing a sample of stars in the
Orion Nebula Cluster found that $\eta$ varies with the stellar mass,
reaching values lower than 1 for $M_\ast \gtrsim 0.5 M_\odot$.  They
also obtained that the exponent $b$ varies with age, and goes from
1.15 for t $\sim$0.8 Myr to 2.43 for t $\sim$10 Myr.
\citetads{2012MNRAS.421...78S} found $\eta = 0.3$ and b = 0.82 in
star forming regions of the Large Magellanic Cloud while
\citetads{2011A&A...525A..47R} found b = 1.6 for the $\sigma$ Ori
region.  \citetads{2014A&A...570A..82V} analyzing the accretion and
variability of stars in NGC 2264 obtained b = 1.4 $\pm$ 0.3.  There
is some controversy also related to the period - mass relation, i.
e., the scarcity of slow rotators among lower mass stars (M$_\ast
< 0.5$ M$_\odot$) which is seen, for example, by
\citetads{2012ApJ...747...51H} in the period - mass diagrams of
several clusters. However, neither \citetads{2013MNRAS.430.1433A}
for NGC 2264 nor \citetads{2013A&A...560A..13M} for h Persei find
any evidence of this relation above 0.4 M$_\odot$.

In this work, we investigate the main variables that can influence
the spin rate evolution of a cluster of stars using Monte Carlo
simulations and compare the results to observations available
in the literature. We randomly assign a mass, a mass accretion
rate and an initial period for each star. We control the disk
lifetime assuming a mass accretion rate threshold below which the
star is considered to be diskless. We run our tests from 1 Myr to
12.1 Myr.  In section (\ref{method}), we explain the main assumptions
of the simulations. In section \ref{results}, we present and discuss
our results. In section \ref{conclusions}, we draw our conclusions.

\section{The Monte Carlo simulations \label{method}}

\begin{figure}
\centerline{\includegraphics[width=8.3cm]{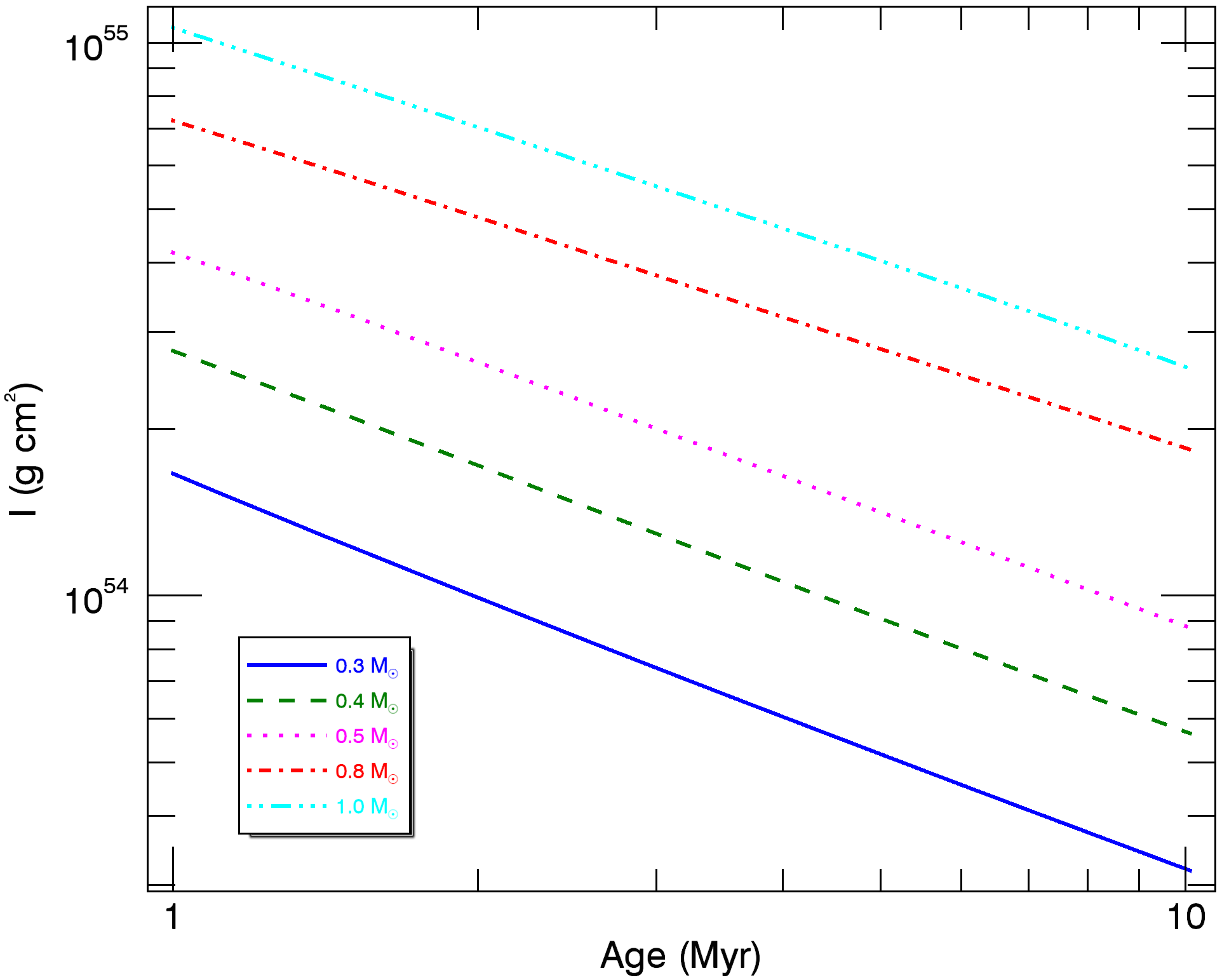}}
\caption{Moments of inertia of 0.3, 0.4, 0.5, 0.8, and 1.0 M$_\odot$
stars as a function  of time.}
\label{momin}
\end{figure}

In order to calculate the evolution of the rotational period $P$,
we assume that a disk star has a constant period and a diskless
star changes its angular velocity conserving angular momentum
following the equation,

\begin{equation} \label{wevol}
\omega(t+ \Delta t) = \omega(t) \frac{I(t)}{I(t+\Delta t)},
\end{equation}
where $\omega(t) = 2 \pi / \mathrm{P(t)}$ is the angular velocity
of the star at time t and $I(t)$ is its moment of inertia at the
same age.  We calculate the moments of the inertia from the stellar
evolutionary models of \citetads{1998A&A...337..403B} for 5 mass
bins: 0.3 M$_\odot$, 0.4 M$_\odot$, 0.5 M$_\odot$, 0.8 M$_\odot$
and 1.0 M$_\odot$.  Figure \ref{momin} shows the interpolated moments
of inertia for each mass bin as a function of the time interval
considered in these simulations.  As expected, the moment of inertia
decreases with time since the star is contracting.  This will
increase the stellar rotation rate and decrease the period of
rotation if the star is free from its disk.

\begin{figure}
\resizebox{\hsize}{!}{\includegraphics{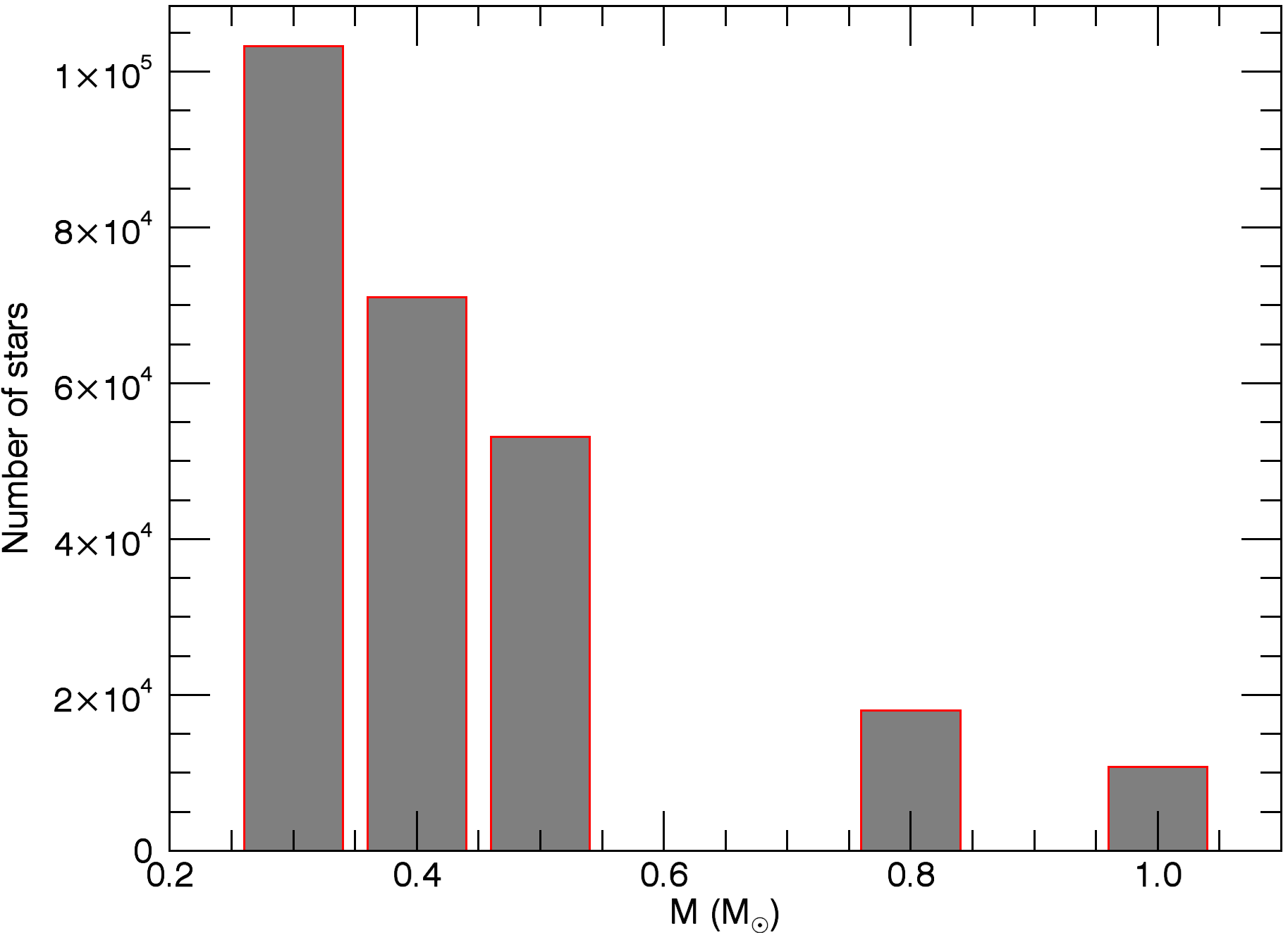}}
\caption{Initial mass distribution for the 5 mass bins considered in this
work.}
\label{massIdist}
\end{figure}

We generate period, mass and mass accretion rate distributions.
Then, a population of about 280,000 stars is evolved from 1 Myr to
12.1 Myr in time steps of 0.1 Myr. The number of stars in each
mass bin is obtained following the canonical IMF
\citepads{2013pss5.book..115K},

\begin{equation} \label{imf}
\xi (m) =
\begin{cases}
\mathrm{k \left(\frac{m}{0.07}\right)^{-1.3}} & \text{$0.07 <$ m
$\leq 0.5$,} \\
\mathrm{k \left[\left(\frac{0.5}{0.07}\right)^{-1.3}\right]
\left(\frac{m}{0.5}\right)^{-2.3}} & \text{0.5 $<$ m $\leq 150$,}
\end{cases}
\end{equation}
where $\mathrm{m}$ is the mass in solar mass units (Figure
\ref{massIdist}). The exact shape of the IMF does not impact
 our results.

In order to obtain the initial mass accretion rate for each star,
we calculate 5 different values, one for each mass bin, according
to the equation:

\begin{equation} \label{Macci}
\dot{M}_\mathrm{acc,i} = \langle \dot{M}_\mathrm{acc} (1 \; \mathrm{Myr}, 
M_\ast) \rangle = c \, M_\ast^b,
\end{equation}
where $M_\ast$ is the value of the stellar mass and b = 1.4
\citepads{2014A&A...570A..82V}. The constant c is obtained considering
that a 1 M$_\odot$ has a mass accretion rate equal to $1.0 \times
10^{-8}$ M$_\odot$ yr$^{-1}$ at t = 1.0 Myr. The values are shown
in Table \ref{tabiMdot}. We then create 5 random log-normal
distributions with means given by $\dot{M}_\mathrm{acc,i}$ (see
equation \ref{Macci} and Table \ref{tabiMdot}) and $\sigma_{\log
\dot{M}_\mathrm{acc}} = 0.8$.

\renewcommand{\arraystretch}{1.3} 

\begin{table}[t]
\caption{Mass accretion rate distributions}
\label{tabiMdot}
\centering
\begin{tabular}{ccc}
\hline\hline
Mass & $\dot{M}_\mathrm{acc, i}$ & $\dot{M}_\mathrm{acc, th}$ \\
(M$_\odot$) & \multicolumn{2}{c}{$(\times \; 10^{-9}$ M$_\odot$ yr$^{-1})$} \\
\hline
0.3 & 1.8 & 0.57 \\
0.4 & 2.8 & 0.85 \\
0.5 & 3.8 & 1.2 \\
0.8 & 7.3 & 2.2 \\
1.0 & 10 & 3.1 \\
\hline
\end{tabular}
\end{table}

\begin{figure}
\resizebox{\hsize}{!}{\includegraphics{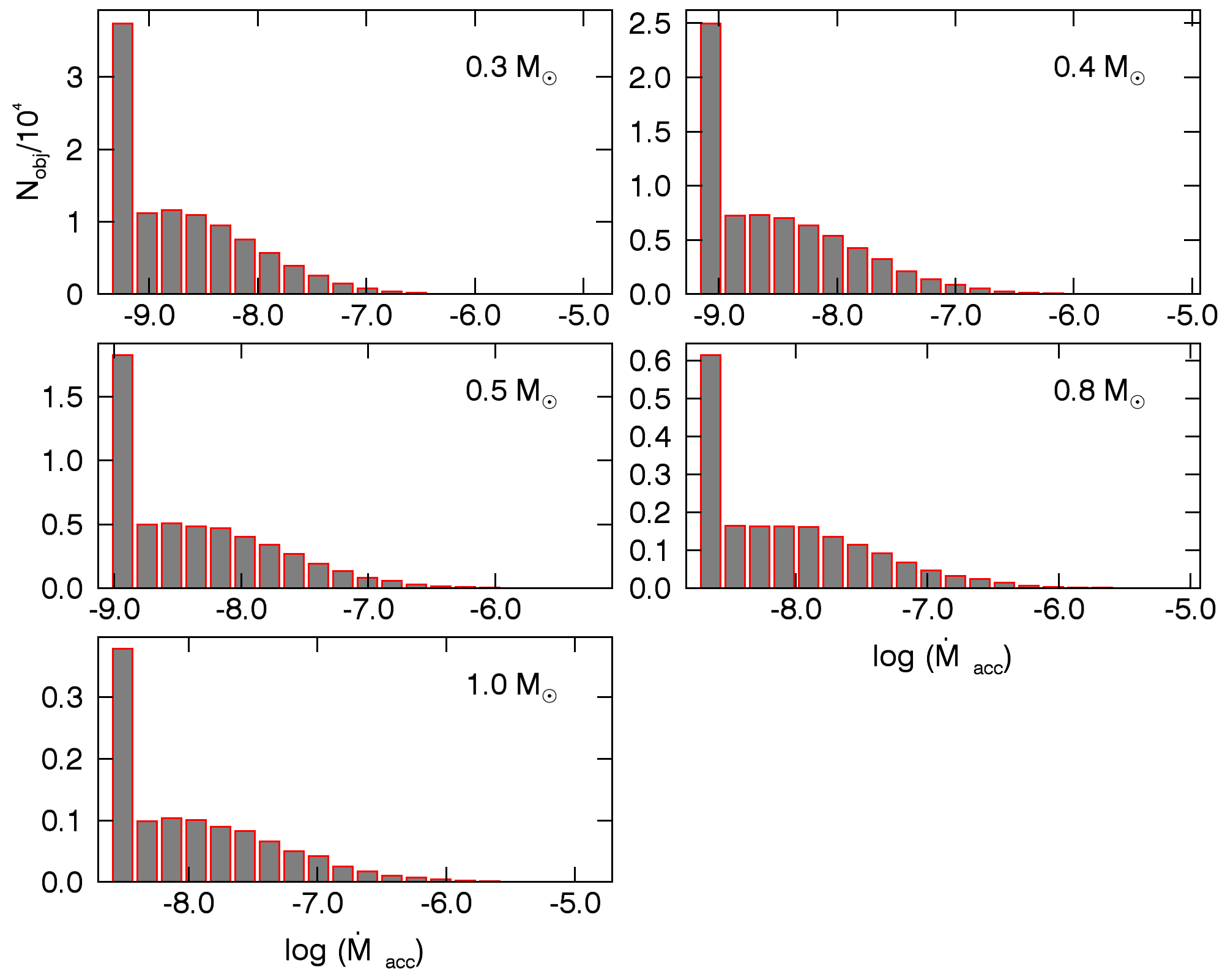}}
\caption{Initial distributions of the mass accretion rate in each
mass bin.}
\label{maccIdist}
\end{figure}

With this initial setup, the mass accretion rate of each star evolves
following the equation,

\begin{equation} \label{Macct}
\dot{M}_\mathrm{acc}(t, M_\ast) = \dot{M}_\mathrm{acc}(1 \; \mathrm{Myr}, 
M_\ast) \, t_{Myr}^{-\eta},
\end{equation}
where the time $t_\mathrm{Myr}$ is expressed in Myr and $\eta =
1.5$ \citepads{1998ApJ...495..385H}. In this equation,
$\dot{M}_\mathrm{acc}(1 \; \mathrm{Myr}, M_\ast)$ is the initial
value of the mass accretion rate of a given star (at 1 Myr). We
then choose a criterion through which the stars will loose their
disks. If, for a given star at a given age, $\dot{M}_\mathrm{acc}
(t, M_\ast) \leq \dot{M}_\mathrm{acc, th} (M_\ast)$, the star will
be released from its disk and will start to spin up conserving
angular momentum. Otherwise it will keep its disk and its rotational
period will be held constant.  The threshold mass accretion rate,
$\dot{M}_\mathrm{acc, th}$, is obtained taking equation (\ref{Macct})
at $t_\mathrm{Myr} = t_\mathrm{th}$, the threshold time, which is
chosen in order to reproduce the observational disk fraction as a
function of age \citepads[e.g.,][]{2009AIPC.1158....3M}.  We take
t$_\mathrm{th}$ = 2.2 Myr, which means that 50\% of the stars will
have lost their disk by this age.  The mass accretion threshold
values for each mass bin are shown in Table \ref{tabiMdot}.  All
mass accretion rate values, including the initial ones, falling
below the threshold are set equal to $\dot{M}_\mathrm{acc, th}$,
which causes the appearance of a peak at this value in the initial
distributions shown in Figure \ref{maccIdist}. Initially diskless
stars start to spin up according to equation (\ref{wevol}).  As
times goes by, different stars will reach the mass accretion rate
threshold, will be released from their disks and will start to spin
up according to the same equation.

\begin{figure*}
\centerline{\includegraphics[width=15cm]{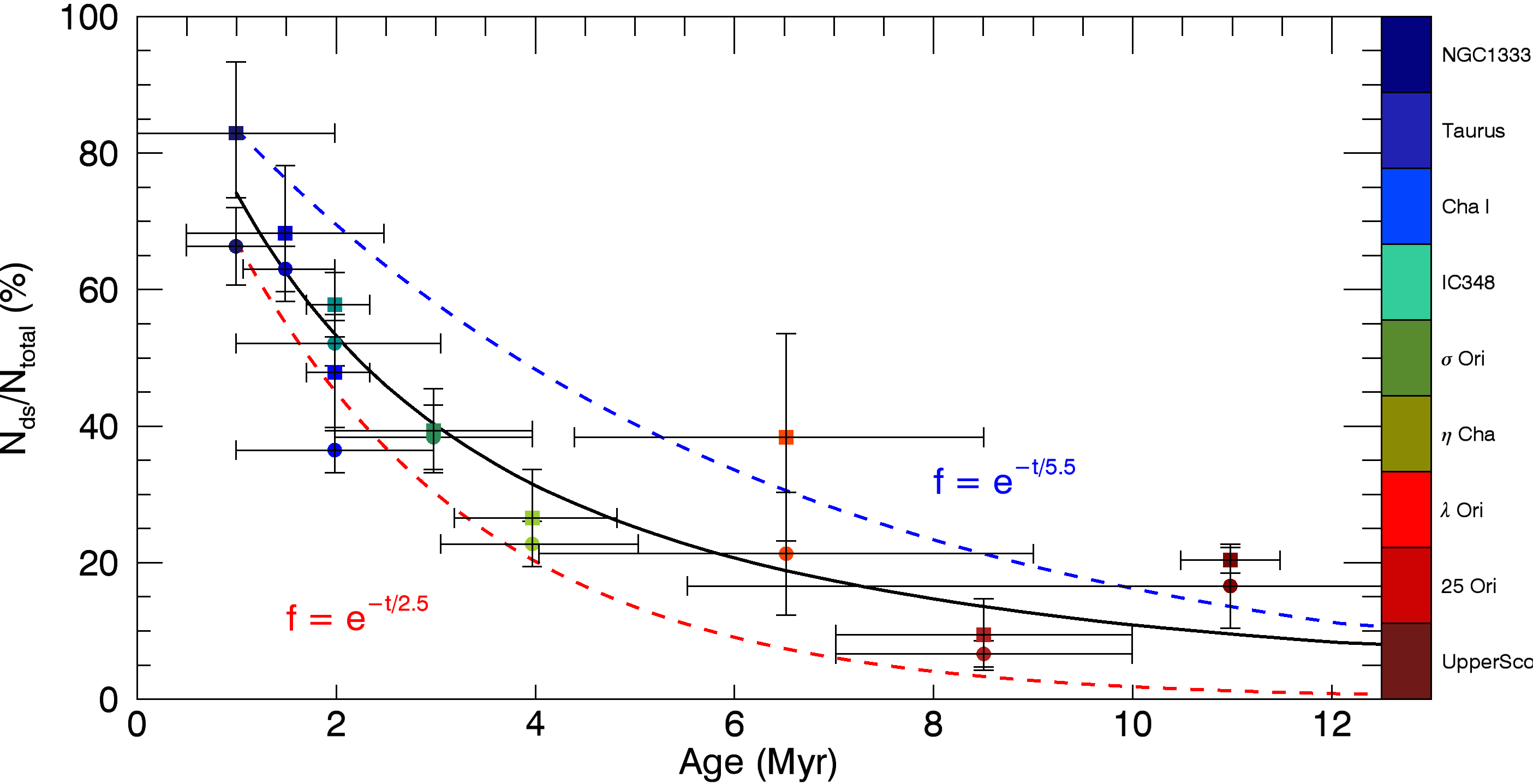}}
\bigskip
\caption{Disk fraction as a function of time (solid black line)
obtained from our simulations considering $\eta = 1.5$ and
t$_\mathrm{th} = 2.2$ Myr. The colored dashed lines are the exponential
decay laws expected from disk e-folding times 2.5 Myr (red line)
and 5.5 Myr (blue line). The colored dots are data for 9 young
nearby associations from \citetads[][circles]{2014A&A...561A..54R},
\citetads{2007ApJ...662.1067H} and
\citetads[][squares]{2008ApJ...686.1195H}.}
\label{diskfrac}
\end{figure*}

In Figure \ref{diskfrac}, we show the disk fraction as a function
of age obtained from our simulations superimposed on data from
\citetads{2014A&A...561A..54R}, \citetads{2007ApJ...662.1067H}, and
\citetads{2008ApJ...686.1195H} for 9 young nearby associations.
The combination of model parameters makes the number of diskless
stars approximately equal to 25\% initially. Our simulated disk
fraction describes well the observed one and it stays between the
two exponential limiting curves based on the short disk lifetimes
claimed by, for example, \citetads{2007AJ....133.2072D} and the
longer disk lifetimes proposed by \citetads{2013MNRAS.434..806B}.
At 12 Myr, the disk fraction is around 8\%.

\section{Results and discussion} \label{results}

We will consider two models M1 and M2 which differ from each
other by the initial distribution of periods of disk and diskless
stars.  Both have b = 1.4, $\eta$ = 1.5 (see equations \ref{Macci}
and \ref{Macct}), t$_\mathrm{th} = 2.2$ Myr and the same mass and
initial mass accretion rate distributions. The resulting disk
fraction (shown in Figure \ref{diskfrac}) will then be the same for
both models since it only depends on the choice of the threshold
time, $\eta$ and the mass accretion rate distribution.

Changing the mass dependency of the mass accretion rate in our
models does not change the main results we obtain below: the disk
fraction, the period distributions, the time dependency of the
median specific angular momentum remain the same. Of course, it
changes the distribution of mass accretion rates. The values are
higher for a model with, for example, b = 0.82, as observed by
\citetads{2012MNRAS.421...78S} at some star forming regions of the
Large Magellanic Cloud. However, the mass accretion rate as a
function of time (equation \ref{Macct}) and the mass accretion rate
threshold also depend on the mass in the same way, so all the problem
in this sense is scale free.  Thus, our models cannot be used to
determine the mass dependency of the mass accretion rate, although
they show a strong dependency on $\eta$ and the mass accretion rate
dispersion ($\sigma_{\log \dot{M}_{acc}}$).  If the mass accretion
rate dispersion is smaller than what we have considered in this
work, the time dependency must change in order to reproduce the
observed disk fraction as a function of time.  For example, if the
mass accretion rate dispersion is less than 2 orders of magnitude,
i.e, narrower than what is considered in this paper, $\eta$ can be
as low as 0.3. This happens because there will be less stars with
high mass accretion rate values, and then the disk fraction will
fall off very quickly.  In order to compensate this, the mass
accretion rate must decrease slowly with time. To have a time
dependency which is in agreement with the self-similar accretion
theory \citepads[i. e.  $\eta > 1$;][]{1998ApJ...495..385H}, we
should have a great amplitude of mass accretion rate values as is
observed in several young clusters (\citeads{2014A&A...570A..82V};
\citeads{2012ApJ...755..154M}).

\subsection{Model M1}

\begin{figure*}
\begin{center}
\includegraphics[width=0.45\textwidth]{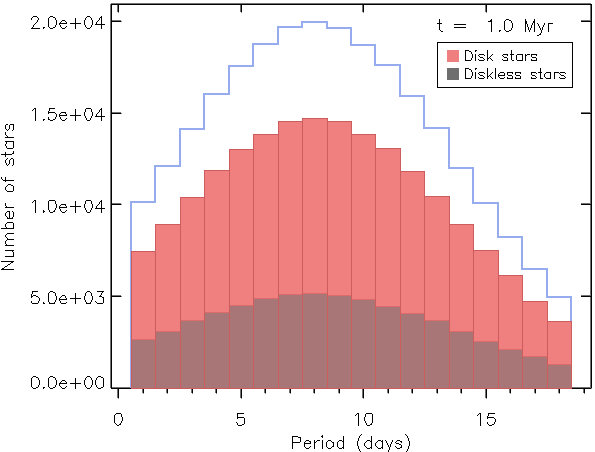}
\hspace{-0.2cm}
\includegraphics[width=0.45\textwidth]{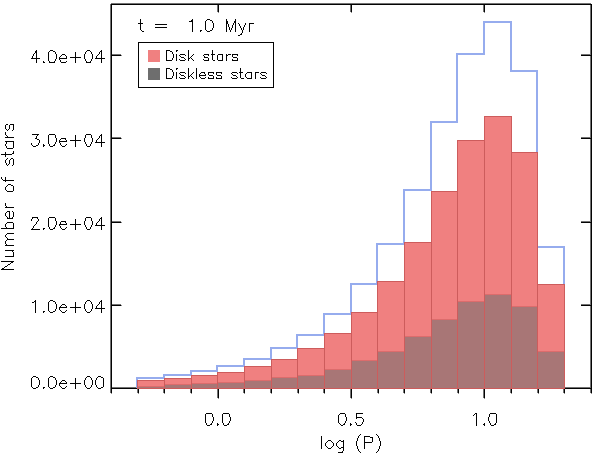}
\includegraphics[width=0.45\textwidth]{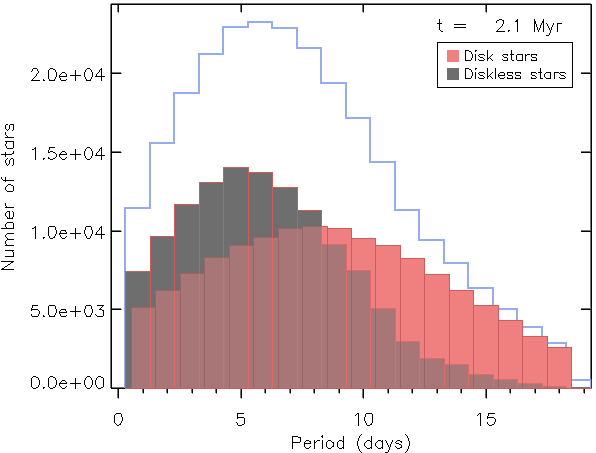}
\hspace{-0.2cm}
\includegraphics[width=0.45\textwidth]{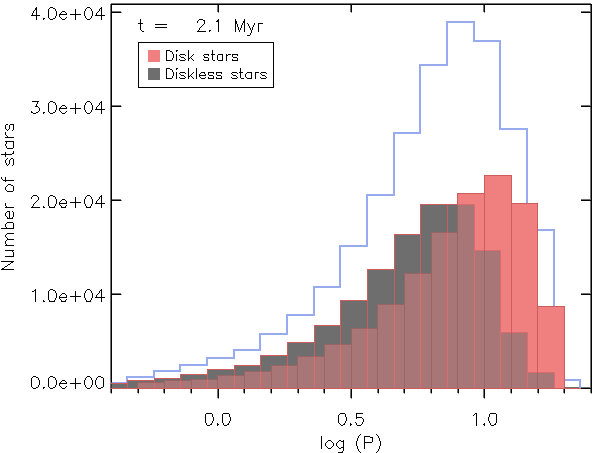}
\includegraphics[width=0.45\textwidth]{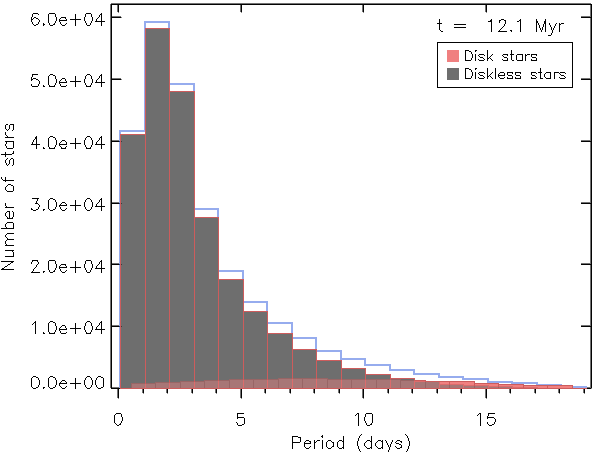}
\hspace{-0.2cm}
\includegraphics[width=0.45\textwidth]{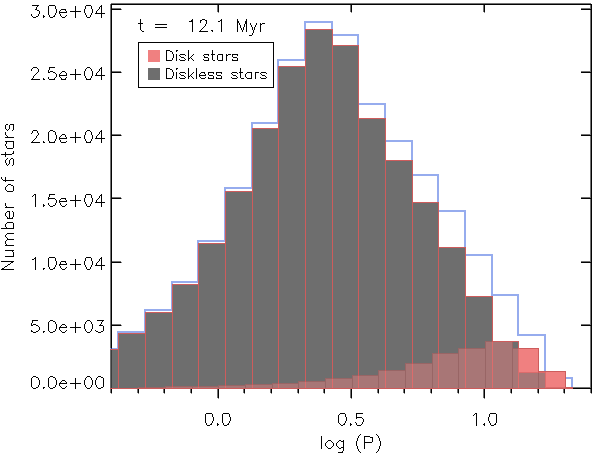}
\end{center}
\caption{Period distributions obtained from model M1 for disk (red)
and diskless (gray) stars. Outlined in blue is the period
distribution for all (disk + diskless) stars.  On top panels are
shown the distributions at t = 1 Myr; central panels at 2.1 Myr;
and bottom panels at 12.1 Myr.  On the left, the period distributions
are shown on a linear scale while on the right, the abscissa is in
log units.  \label{distP}}
\end{figure*}

In model M1, we consider that the initial period distribution
is the same for disk and diskless stars. It is given by a Gaussian
function with a mean period equal to 8 days and a standard deviation
equal to 6 days (e.  g., \citeads{2004AJ....127.1029R};
\citeads{2014prpl.conf..433B}). The Gaussian is truncated at both
ends, at 0.5 days, and at 18.5 days. In Figure \ref{distP} we show
the distributions of the rotational periods at t = 1.0, 2.1, and
12.1 Myr.  At t = 1.0 Myr, our model assumptions establish an initial
random distribution of periods and mass accretion rates for the
whole population.  According to the initial fraction of diskless
stars derived from Figure \ref{diskfrac} (equal to 26\%), the number
of diskless stars is smaller than the number of disk stars.  This
changes at t = 2.1 Myr, with the fraction of diskless stars increasing,
and the distributions now peak at P$_\mathrm{peak} = 8 - 9$ days
for disk stars and at P$_\mathrm{peak} = 4 - 6$ days for diskless
stars. Our simulations point to a significant overlap between the
rotational distributions of disk and diskless stars at all ages.
Diskless stars tend to rotate faster than disk ones but exhibit a
long tail towards long periods. Conversely, a fraction of fast
rotators is found among disk stars.

\begin{figure*}
\begin{center}
\includegraphics[width=0.45\textwidth]{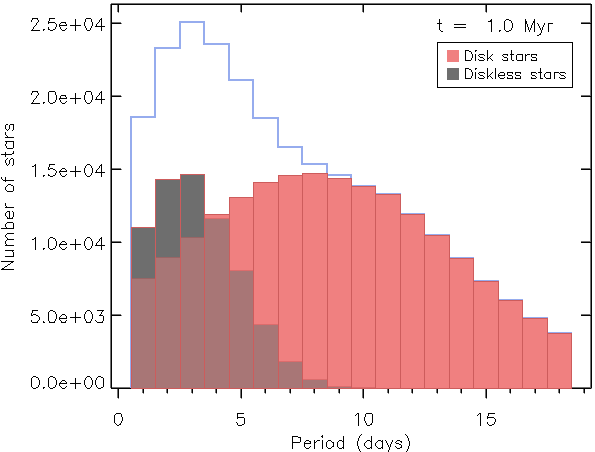}
\hspace{-0.2cm}
\includegraphics[width=0.45\textwidth]{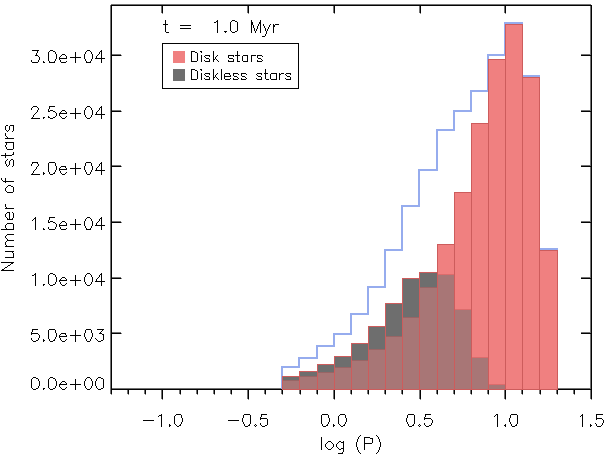}
\includegraphics[width=0.45\textwidth]{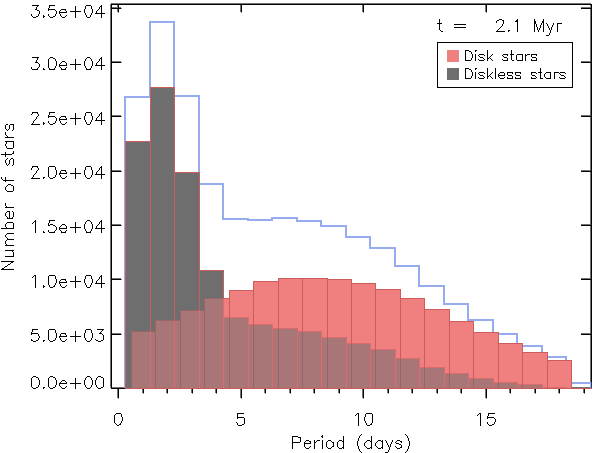}
\hspace{-0.2cm}
\includegraphics[width=0.45\textwidth]{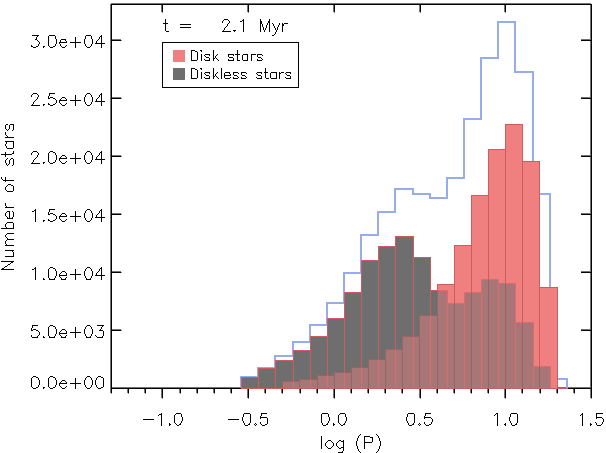}
\includegraphics[width=0.45\textwidth]{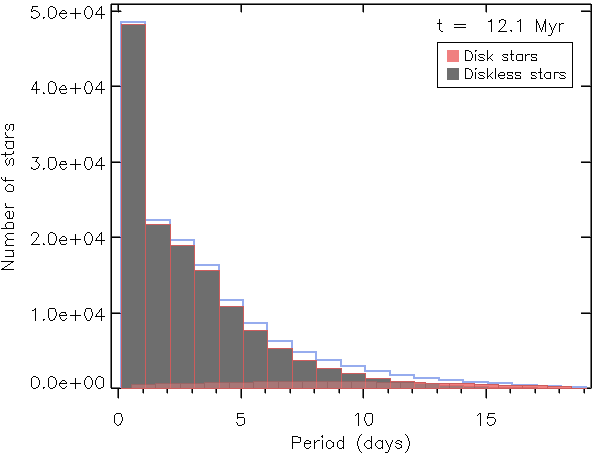}
\hspace{-0.2cm}
\includegraphics[width=0.45\textwidth]{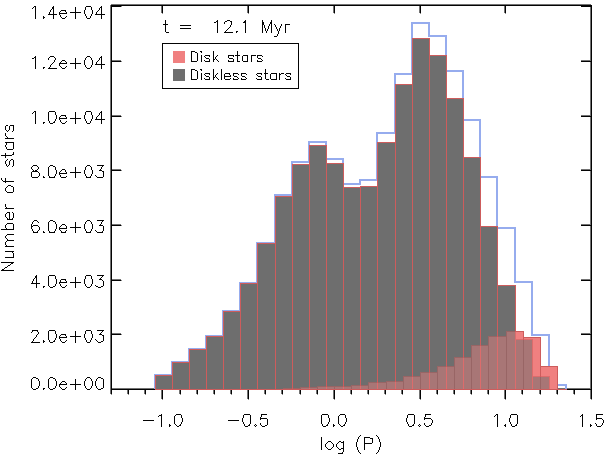}
\end{center}
\caption{Period distributions obtained from model M2 for disk (red)
and diskless (gray) stars. Outlined in blue is the period
distribution for all (disk + diskless) stars.  On top panels are
shown the distributions at t = 1 Myr; central panels at 2.1 Myr;
and bottom panels at 12.1 Myr.  On the left, the period distribution
are shown on a linear scale while at right, the abscissa is in log
units.  \label{distPM2}}
\end{figure*}

\citetads{2012ApJ...747...51H} analyzing a sample of stars in NGC
6530 (1 - 2 Myr) find that the mean period of stars showing NIR
excess is around 6.3 days and $\bar{\mathrm{P}} = 3.7$ days for
stars without NIR excess.  \citetads{2013MNRAS.430.1433A} also find
$\bar{\mathrm{P}} = 7$ days for CTTS and $\bar{\mathrm{P}} = 4.2$
days for WTTS in a sample of stars in NGC 2264 (3 Myr).
\citetads{2007ApJ...671..605C} found that stars with a spectral
type M2 and earlier with and without disks also present different
period distributions for NGC 2264 (1 - 5 days for diskless stars
and a flatter distribution for disk stars) and the ONC (peak at
$\sim 2$ days for diskless and at $\sim 8$ days for disk stars).
Our results are in reasonable agreement with these observations.
However, at 12.1 Myr, our simulation does not reproduce the bimodal
rotational distribution of h Per cluster members, which exhibit one
peak at $\leq 1$ day and another at 3 - 7 days
\citepads{2013A&A...560A..13M}.  Instead, in our simulation, the
period distribution of diskless stars moves to smaller periods as
a whole by an age of 12.1 Myr with a single peak at P$_\mathrm{peak}
\sim 2$ days. This suggest that the initial conditions of the
simulations might have to be modified.

\subsection{Model M2}

In order to see if we can obtain a bimodal distribution for the
whole population of stars up to 13 Myr we run model M2, for which
the initial period distribution is different for disk and diskless
stars. For disk stars, the period distribution is the same considered
in model M1 (P$_\mathrm{mean} = 8$ days and $\sigma = 6$ days). For
the initial population of diskless stars, the period distribution
is a truncated Gaussian with a peak at 3 days and a dispersion equal
to 2 days. There are no stars with periods less than 0.5 days and
all the stars that originally had periods less than this value are
smoothly redistributed around 3 days. These parameters were
chosen supposing that at 1.0 Myr the stars in a given cluster have
already experienced some disk and rotational evolution. That there
is some disk loss at this age is seen in very young clusters as, for
example, NGC 1333 which is 1 Myr old and has a disk fraction equal
to 66\% $\pm$ 6\% \citepads[][Figure \ref{diskfrac}]{2014A&A...561A..54R}.
Then we suppose that at 1 Myr there is a diskless population which
is spinning up. On the somewhat older ONC cluster,
\citetads{2007ApJ...671..605C} find that stars without signs of the
presence of dust disks rotate faster than the disk stars, with a
mean period of 2 days but with a tail towards longer periods.
Supposing that at 1 Myr these stars rotate slowly than at 2 Myr,
we choose the mean period of the diskless population to be equal
to 3 days but with a dispersion value great enough to allow the
existence of diskless stars with periods as long as 10 days.

\subsubsection{Evolution of the rotational distributions}

In Figure \ref{distPM2} we show the period distribution at 1.0,
2.1, and 12.1 Myr. The distinct initial periods distributions for
disk and diskless stars are seen at 1 Myr. While there are
no diskless stars with periods greater than 10 days, there are still
fast rotators among disk stars. At 2.1 Myr the period distribution
remains clearly bimodal. The peak of the fast rotators is around 2
days and that of the slow rotators is around 8 days as seen by
\citetads{2007ApJ...671..605C} for the ONC. We also recover a bimodal
distribution seen in the distribution of the logarithm of the period
at 12.1 Myr for h Per stars, quite in contrast with model M1 above.
At this age, the peaks of the simulated distribution lie at 0.7 and
3.2 days. While the former is in good agreement with the peak of h
Per fast rotators reported by \citetads{2013A&A...560A..13M}, the
latter is located at smaller periods than observed in h Per where
it lies at about 3-7 days in the mass range 0.4-1.1 M$_\odot$.

\begin{figure}[!t]
\begin{center}
\resizebox{\hsize}{!}{\includegraphics{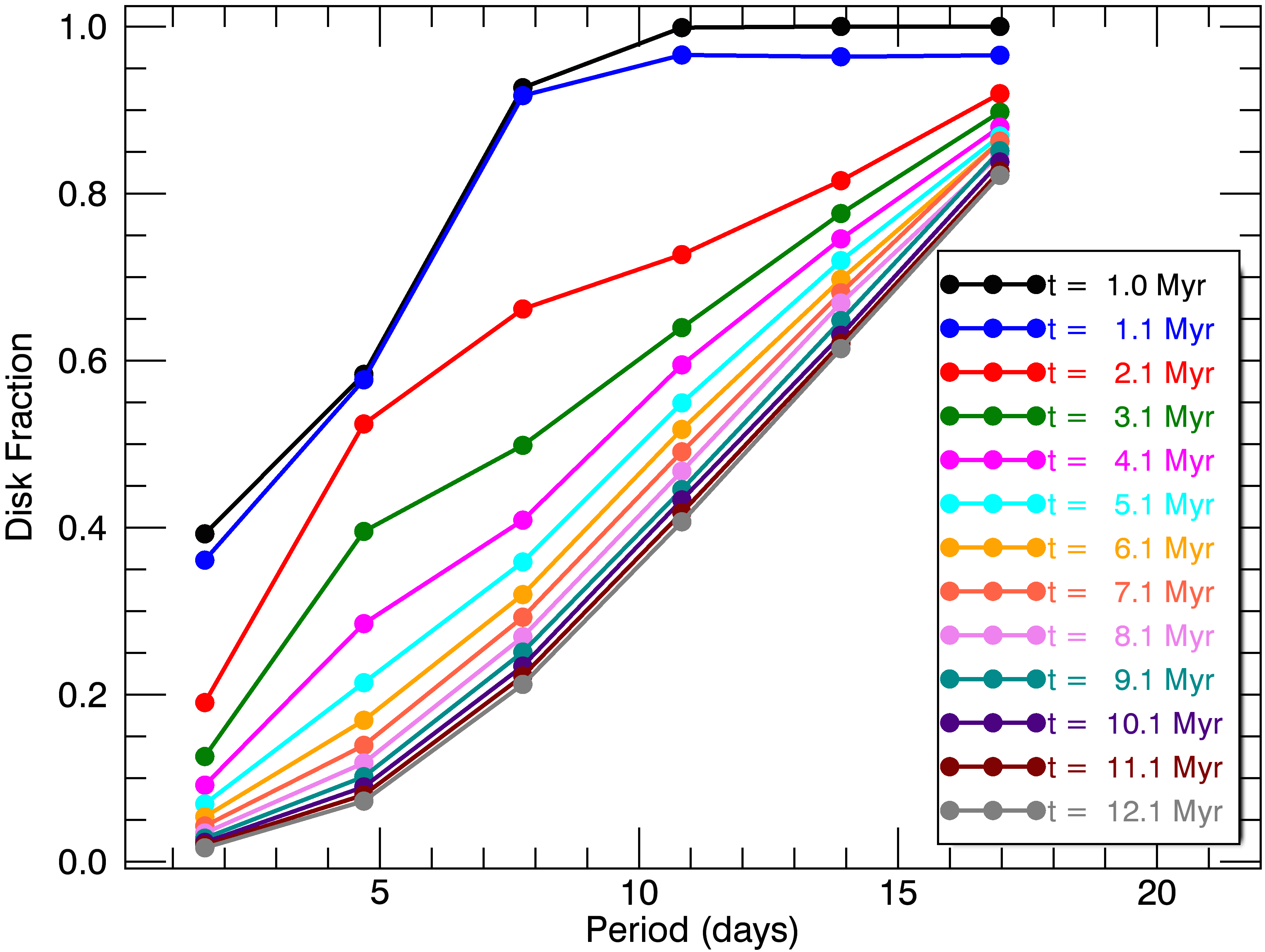}}
\end{center}
\caption{Disk fraction as a function of period for different ages
for model M2.}
\label{diskfracP}
\end{figure}

\subsubsection{Rotation - disk fraction connection}

\citetads{2007ApJ...671..605C} stated in a different way that disk
stars rotate more slowly on average than diskless stars. They showed
that the fraction of disk stars increases with period for a sample
of young stars in NGC 2264.  In Figure \ref{diskfracP}  we plot the
fraction of disk stars as a function of period for model M2 at
different ages. There is a general trend of increasing disk fraction
with increasing period.  At 1 Myr, all stars with periods greater
than 10 days are disk stars (i.e., disk fraction equal to 1.0),
following the M2 initial period distributions. As the system evolves,
the disk fraction as a whole decreases but the decrease rate is
greater at smaller periods. This is due to the fact that slowly
rotating stars that loose their disk start to spin-up, thus leading
to an accumulation of diskless stars at shorter periods.

\begin{figure*}
\begin{center}
\includegraphics[width=8.5cm]{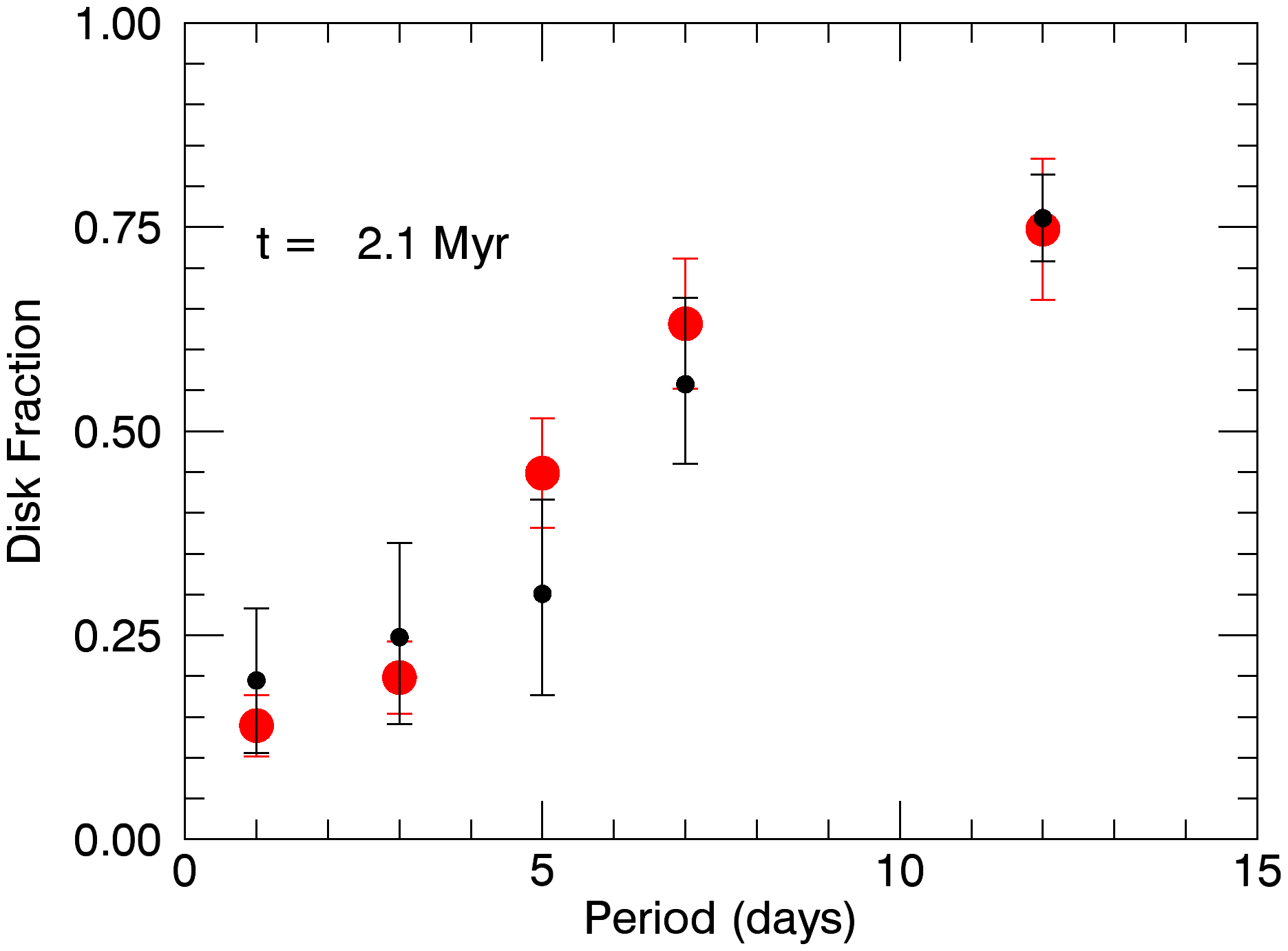}
\includegraphics[width=8.5cm]{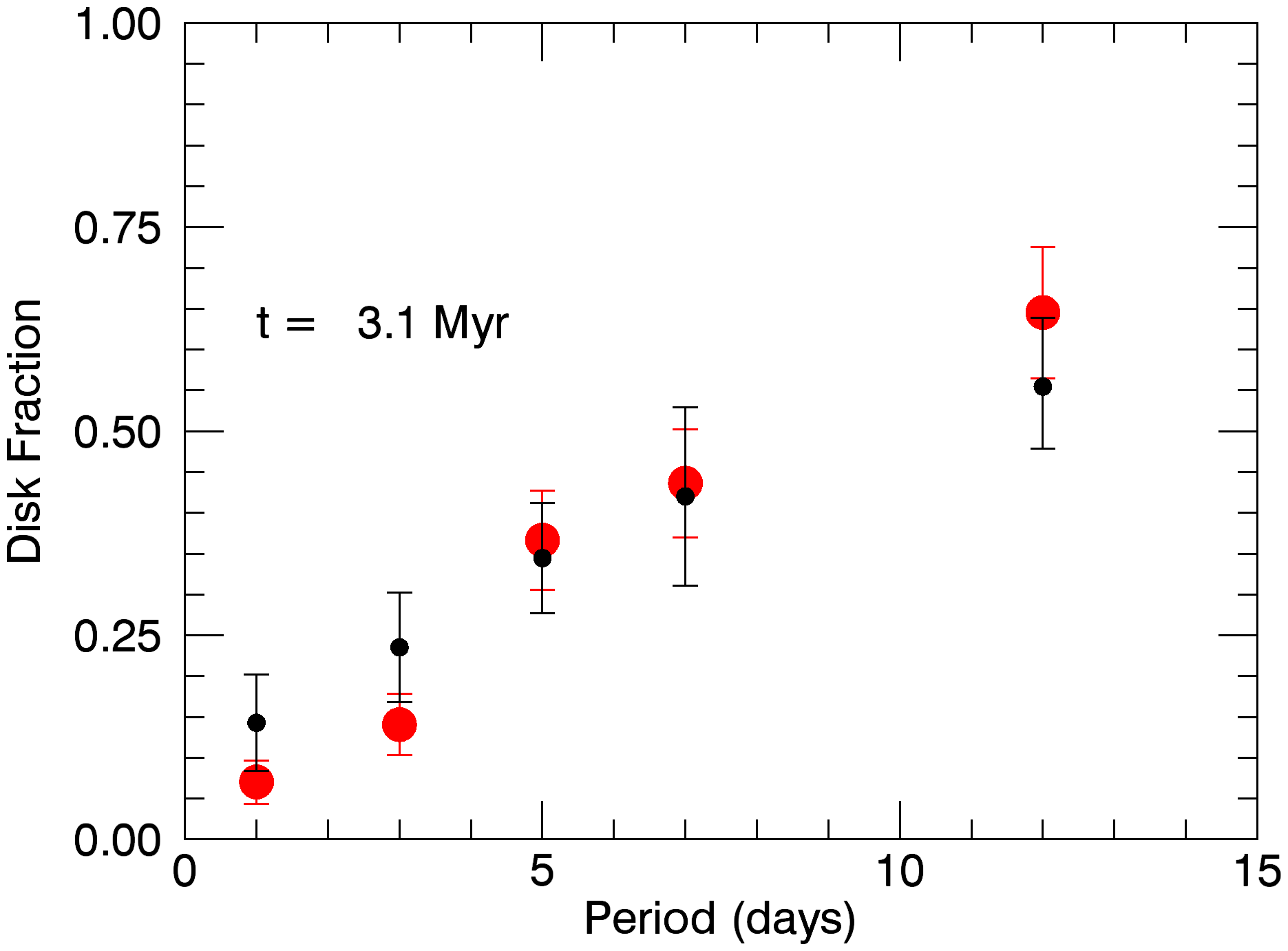}
\end{center}
\caption{Disk fraction as a function of period obtained from model
M2 (red circles) for t = 2.1 Myr (left panel) and for t = 3.1
Myr (right panel). The superimposed black circles show the
disk fraction as a function of the period obtained for the ONC (
left panel) and for NGC 2264 (right panel) for stars
with a spectral type M2 and earlier extracted from
\citetads{2007ApJ...671..605C}.} \label{comparcieza}
\end{figure*}

In Figure \ref{comparcieza} we compare our results at 2.1 and 3.1
Myr  with the data of \citetads{2007ApJ...671..605C} for stars of
ONC and NGC 2264 with a spectral type M2 and earlier.  We use the
same period bins in order to facilitate the comparison.  There is
a good agreement between our results and \citetads{2007ApJ...671..605C}'s
data taking into account the error bars which in our case are equal
to the standard errors of a Poisson counting. The disk fraction is
seen to smoothly increase with rotational period at 2.1 Myr
and 3.1 Myr as observed, which supports the M2 model assumptions,
including disk locking and an initial bimodal distribution of periods
for disk and diskless stars.

\subsubsection{Rotation - accretion connection}

\begin{figure*}[!t]
\begin{center}
\includegraphics[width=0.24\textwidth]{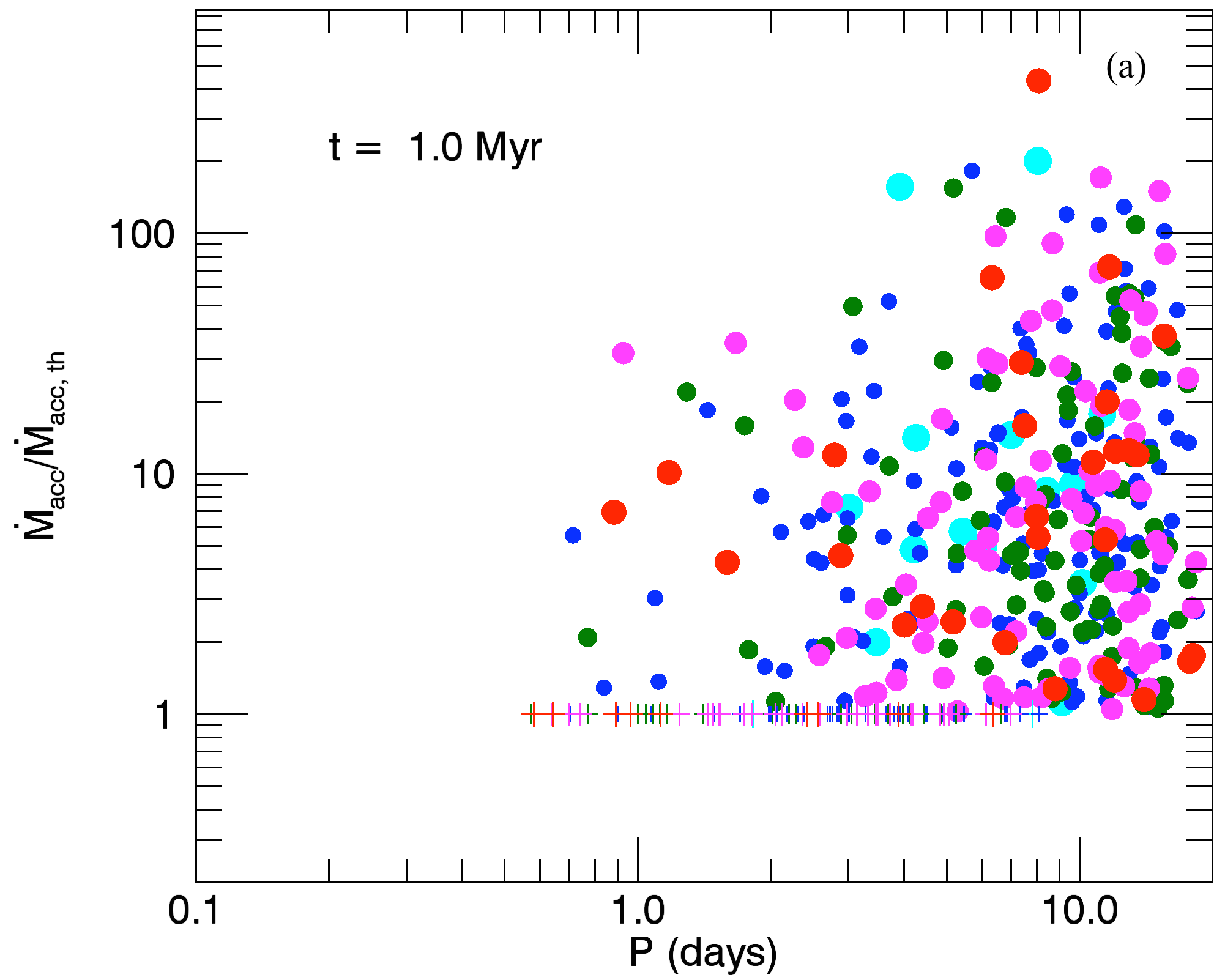}
\hspace{-0.1 cm}
\includegraphics[width=0.24\textwidth]{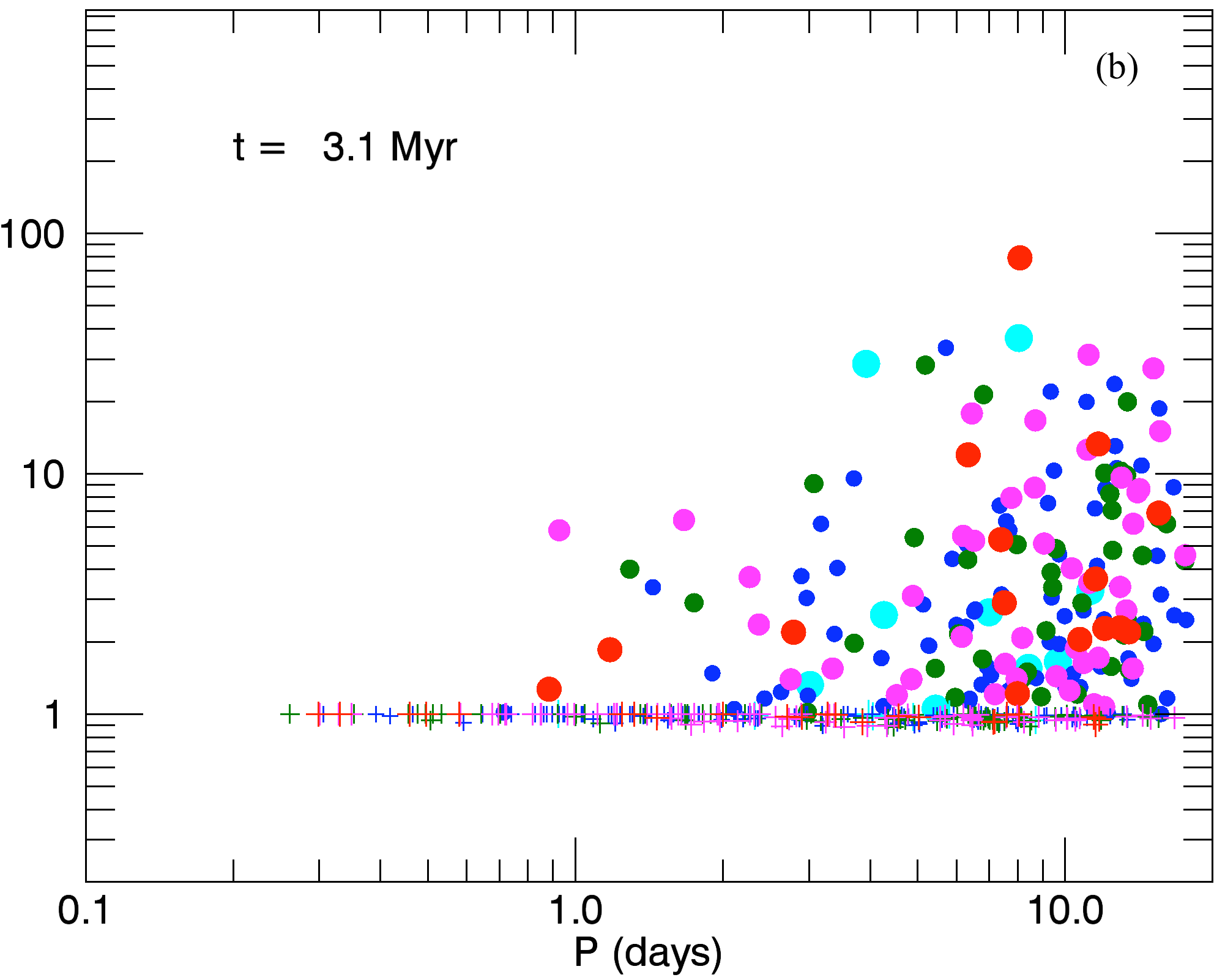}
\vspace{0.2 cm}
\includegraphics[width=0.24\textwidth]{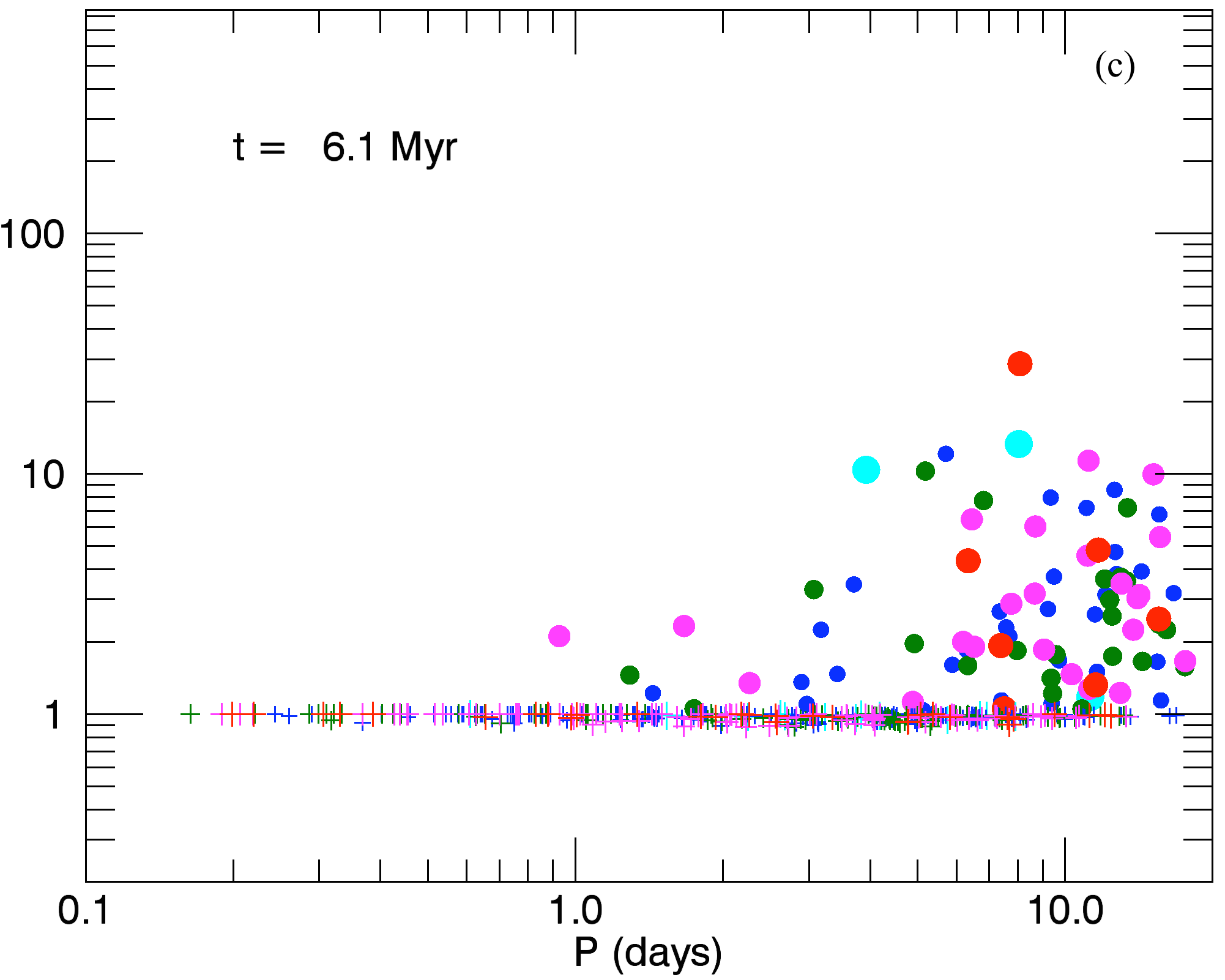}
\hspace{-0.1 cm}
\includegraphics[width=0.24\textwidth]{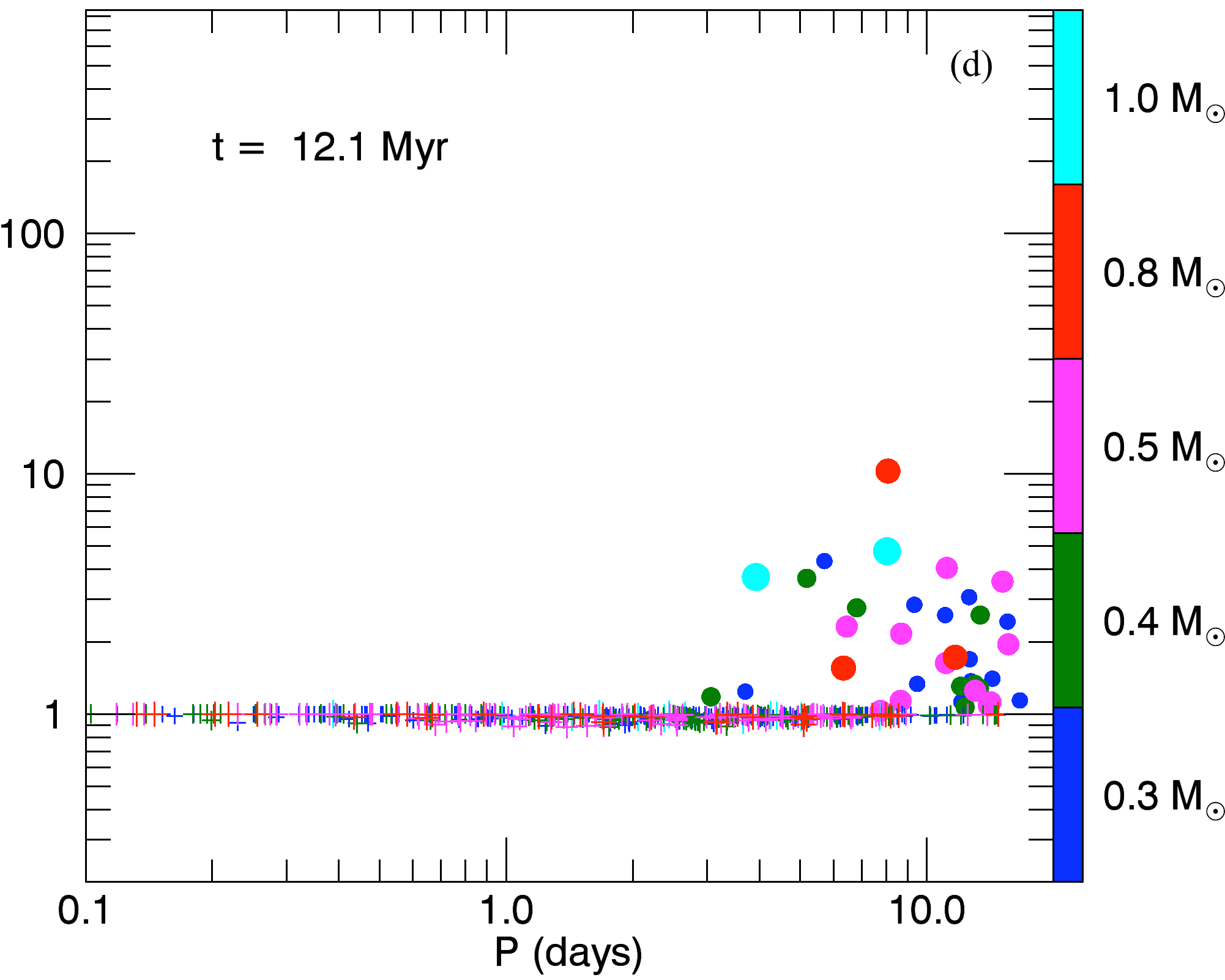}
\end{center}
\caption{$\dot{M}_\mathrm{acc}$ normalized to the mass accretion
rate threshold is plotted versus period at t = 1.0 Myr (panel 
a), t = 3.1 Myr (panel b), t = 6.1 Myr (panel c) and t
= 12.1 Myr (panel d).  For clarity, a subsample of 464 stars
randomly selected from model M2 is shown.  Different colors represent
different mass bins. Circles are for disk stars and crosses are for
diskless stars.  \label{MaccP}}
\end{figure*}

In order to further investigate if we can see a segregation in
period due to the presence of a disk we plot, in Figure \ref{MaccP},
the mass accretion rate normalized to the mass accretion rate
threshold versus rotational period for the simulated sample at an
age of 1.0, 3.1, 6.1, and 12.1 Myr.  At t = 1.0 Myr, diskless stars
$(\dot{M}_\mathrm{acc} \leq \dot{M}_\mathrm{acc, th})$ are set at
$\dot{M}_\mathrm{acc} = \dot{M}_\mathrm{acc, th}$ and none has a
period longer than 10 days. Conversely, there are only few disk
stars at short periods. As the population evolves, the period
distribution of diskless stars gets wider, both towards longer and
shorter periods. The former effect results from the evolution of
disk stars that only recently have lost their disks and that have
not had enough time to significantly spin up, thus yielding a
population of long period diskless stars; the latter is  due to the
free spin up of initially diskless stars as they contract, thus
shifting the bulk of the diskless stars rotational period distribution
towards shorter periods. As a result, after a few Myr, the initially
peaked diskless star period distribution spreads over the full range
of periods from less than 1 day up to 15 days. In contrast,
disk stars tend to fill the locus of longer periods, as they remain
locked to their disk as they evolve. Hence, over time, the initial
correlation between rotational periods and disks remains, but gets
blurred by the widening of the period distribution of diskless
stars. Yet, the short period locus remains depleted of disk stars
at all ages, up to 12.1 Myr.

\begin{figure*}
\begin{center}
\includegraphics[width=0.5\textwidth]{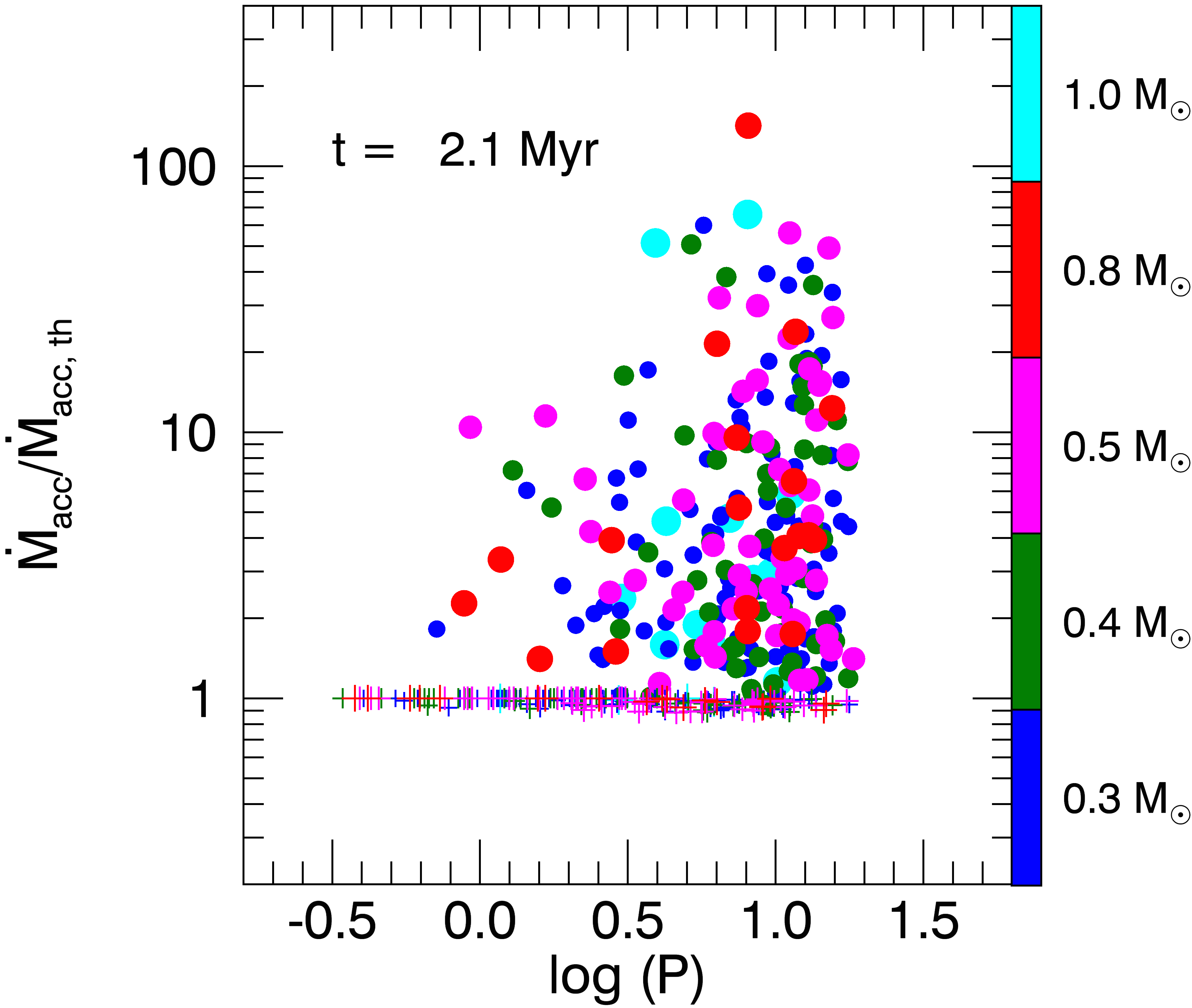}
\hspace{-0.2 cm}
\includegraphics[width=0.41\textwidth]{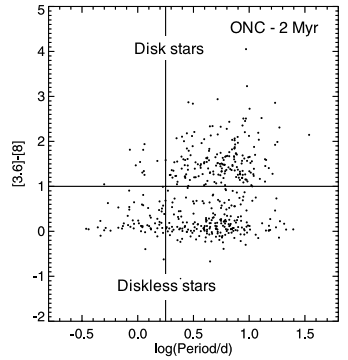}
\end{center}
\caption{{\it Left:} $\dot{M}_\mathrm{acc}$ normalized to the mass
accretion rate threshold  {\it vs.} period at t = 2.1 Myr for a
sample of 464 stars randomly chosen from our M2 model simulation.
Different colors represent different mass bins. Circles are for
disk stars and crosses are for diskless stars. {\it Right:} [3.6]
- [8] IRAC colors for 464 stars with measured periods in the Orion
Nebula Cloud. Figure adapted from \citetads{2006ApJ...646..297R}.
\label{MaccP2}}
\end{figure*}

In Figure \ref{MaccP2}, we compare the simulated $\dot M_\mathrm{acc}$
- $\log P$ plot  at 2.1 Myr with the observed Spitzer IR excess vs.
period plot for ONC from \citetads{2006ApJ...646..297R}. It is not
straightforward to translate mass accretion rate to Spitzer IR
excess, and we therefore did not attempt such a conversion.
Nevertheless, the mid-IR excess is expected to scale, at least in
a statistical sense, with mass accretion rate. In this Figure,
\citetads{2006ApJ...646..297R} show the [3.6] - [8] IRAC colors for
464 stars with measured periods in the Orion Nebula Cloud for which
the mean age is about 2 Myr.  Colors above [3.6] - [8] = 1 are
indicative of the presence of disks. We clearly see in this Figure
a deficit of disk stars at short periods and an accumulation of
diskless stars at all periods. As discussed above, similar trends
are seen at 2.1 Myr in our simulations: the diskless stars cover
the full range of periods and there is a strong deficit of disk
stars at P $ \lesssim 4$ days. Hence, the simulated results of model
M2 yield a good agreement with the observed IR - rotation plots.

\subsubsection{Period - mass relationship}

\begin{figure*}[!t]
\begin{center}
\includegraphics[width=8.5cm]{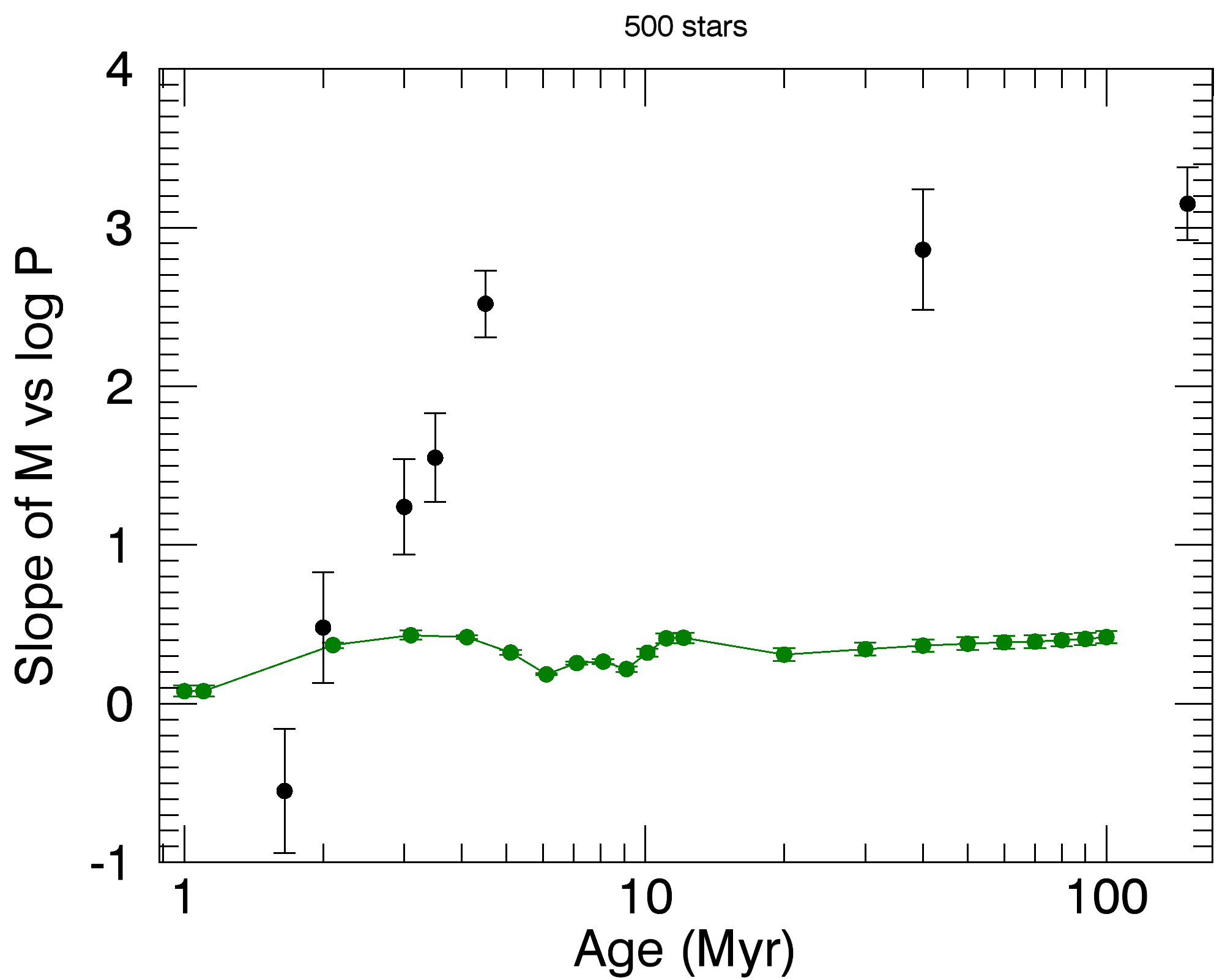}
\includegraphics[width=8.5cm]{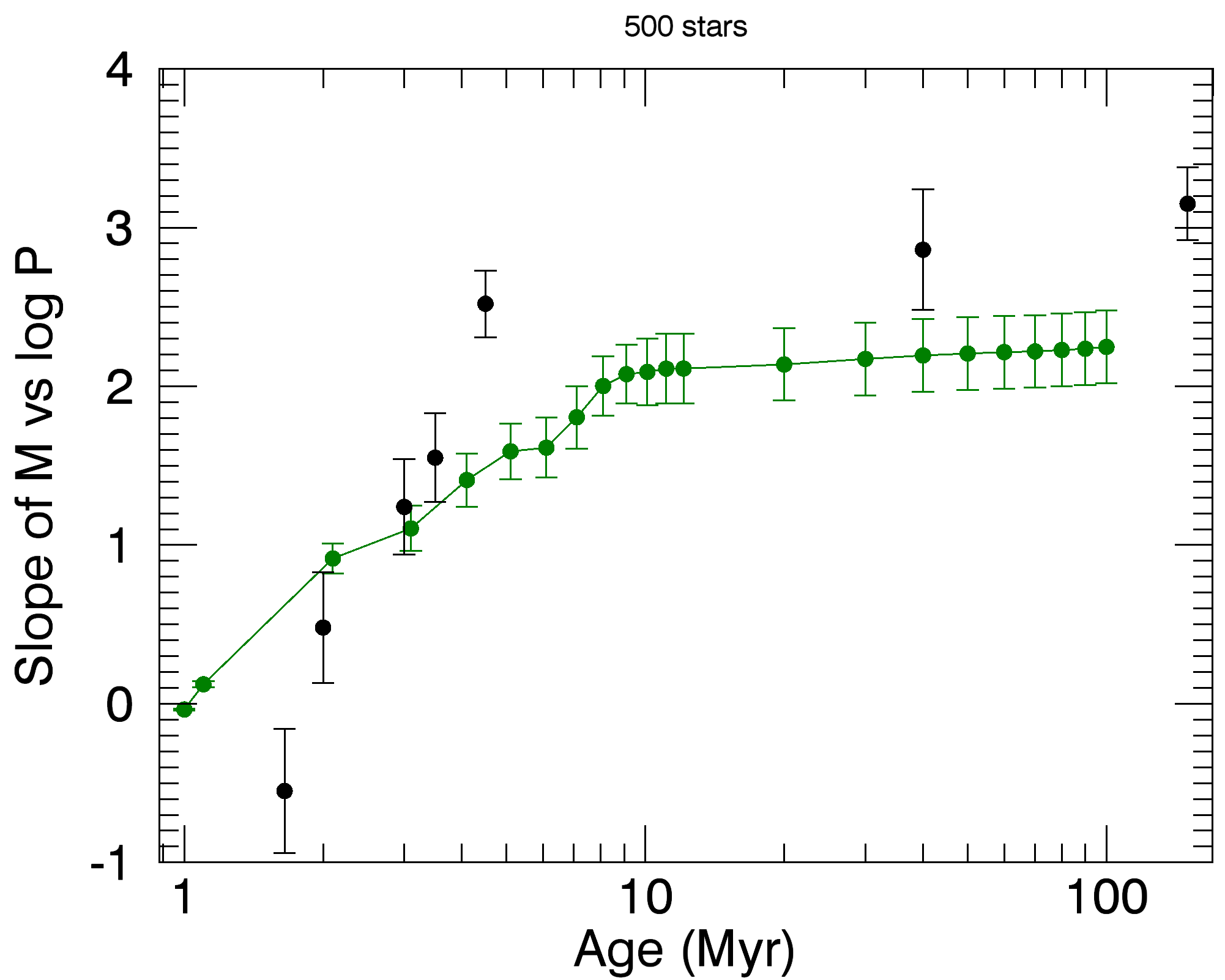}
\end{center}
\caption{{\it Left:} Slope of the $\log$P - mass relationship as a
function of age recovered from the simulations (green dots) and
superimposed to data points from \citetads{2012ApJ...747...51H} for
7 nearby clusters.  The error bars of the green dots are equal to
the standard errors of the estimate of the least-square fits. {\it
Right:} The same plot for a modified model where there is no
disk-locking for the 0.3 M$_\odot$ stars. \label{slopePM}}
\end{figure*}

\citetads{2012ApJ...747...51H} investigated the period - mass
relationship of 7 clusters (NGC 6530, ONC, NGC 2264, NGC 2362, IC
348, NGC 2547 and NGC 2516) with ages ranging from 1 - 2 Myr to 150
Myr. In analyzing low mass (0.2-0.5 M$_\odot$) cluster members they
found a positive correlation for older clusters, meaning that lower
mass stars spin up faster than higher mass ones. Since the mass
range we explore in our simulations does not extend down to 0.2
M$_\odot$, we cannot claim that we can recover such a relationship,
but we could perhaps expect some weaker correlation among the lowest
mass stars of our sample.  Following \citetads{2012ApJ...747...51H},
we plot the slope of the $\log$P - mass relationship  as a
function of age in Figure \ref{slopePM}.  The slopes were
calculated\footnote{Using the LINFIT function of IDL version 8.4
(Exelis Visual Information Solutions, Boulder, Colorado) which is
based in a $\chi^2$ minimization algorithm.} for the $75^\mathrm{th}$
percentile of a random sample of 500 stars in the mass interval
$0.3 \leq$ M/M$_\odot \leq 0.5$ as a function of mass. The slopes
obtained from model M2 do not vary much with age and remain close
to zero. This suggests that the rotational distributions are fairly
independent of mass at very low masses in our simulations, a result
clearly at odds with observations.

We therefore attempted another model with the same assumptions and
parameters as model M2 but relaxing the disk locking assumption for
the lowest mass stars. In the lowest mass bin, 0.3 M$_\odot$, even
a disk star is now assumed to evolve conserving its angular momentum.
The results, shown in Fig.~\ref{slopePM}, yield positively increasing
slopes as a function of time, which qualitatively matches the
observations. This suggests that the disk locking might be less
efficient  at very low-masses, thus allowing these stars to spin
up faster than higher mass ones (cf. \citeads{2005A&A...430.1005L};
\citeads{2014prpl.conf..433B}).

\begin{figure*}
\begin{center}
\includegraphics[width=8.3cm]{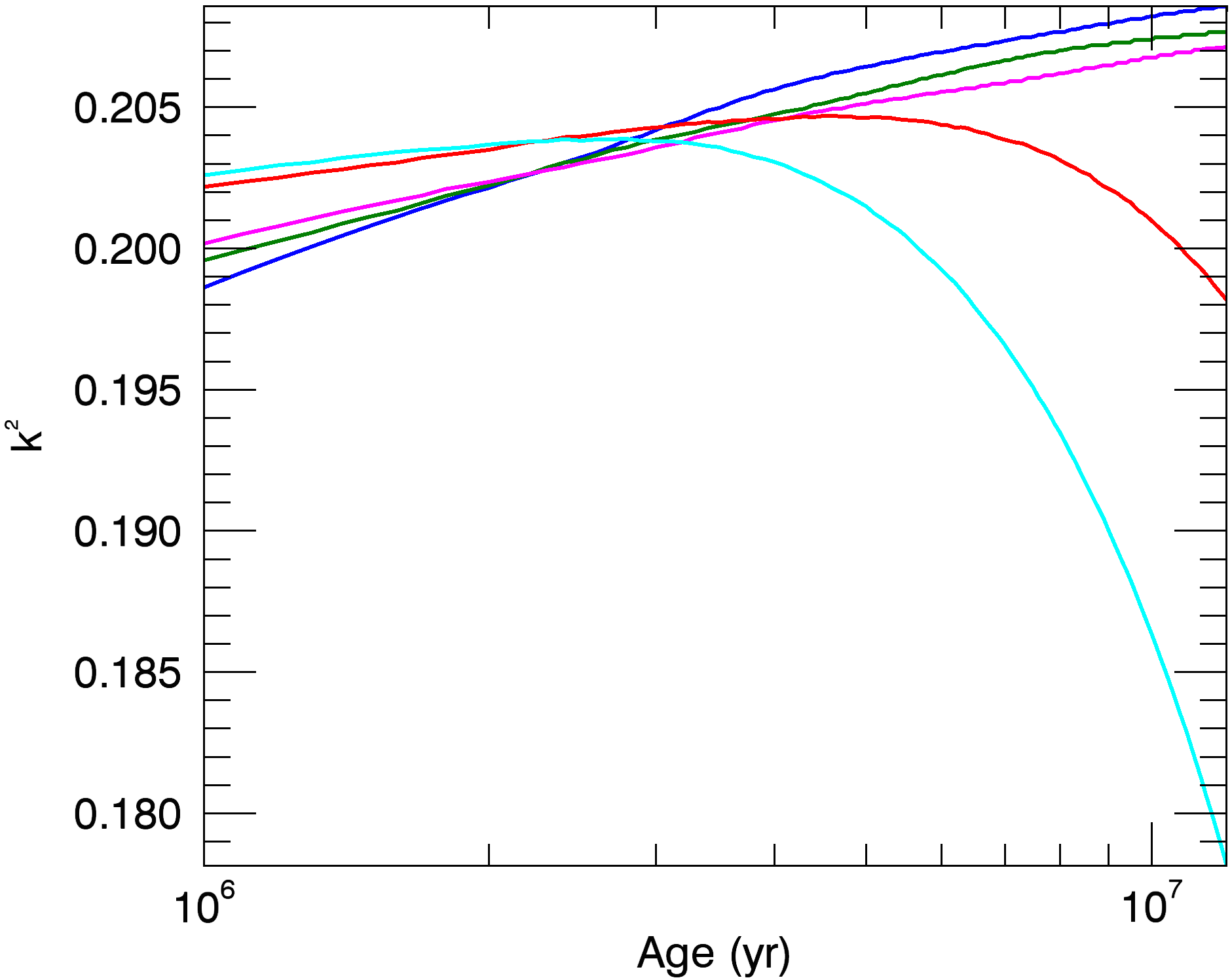}
\includegraphics[width=7.7cm]{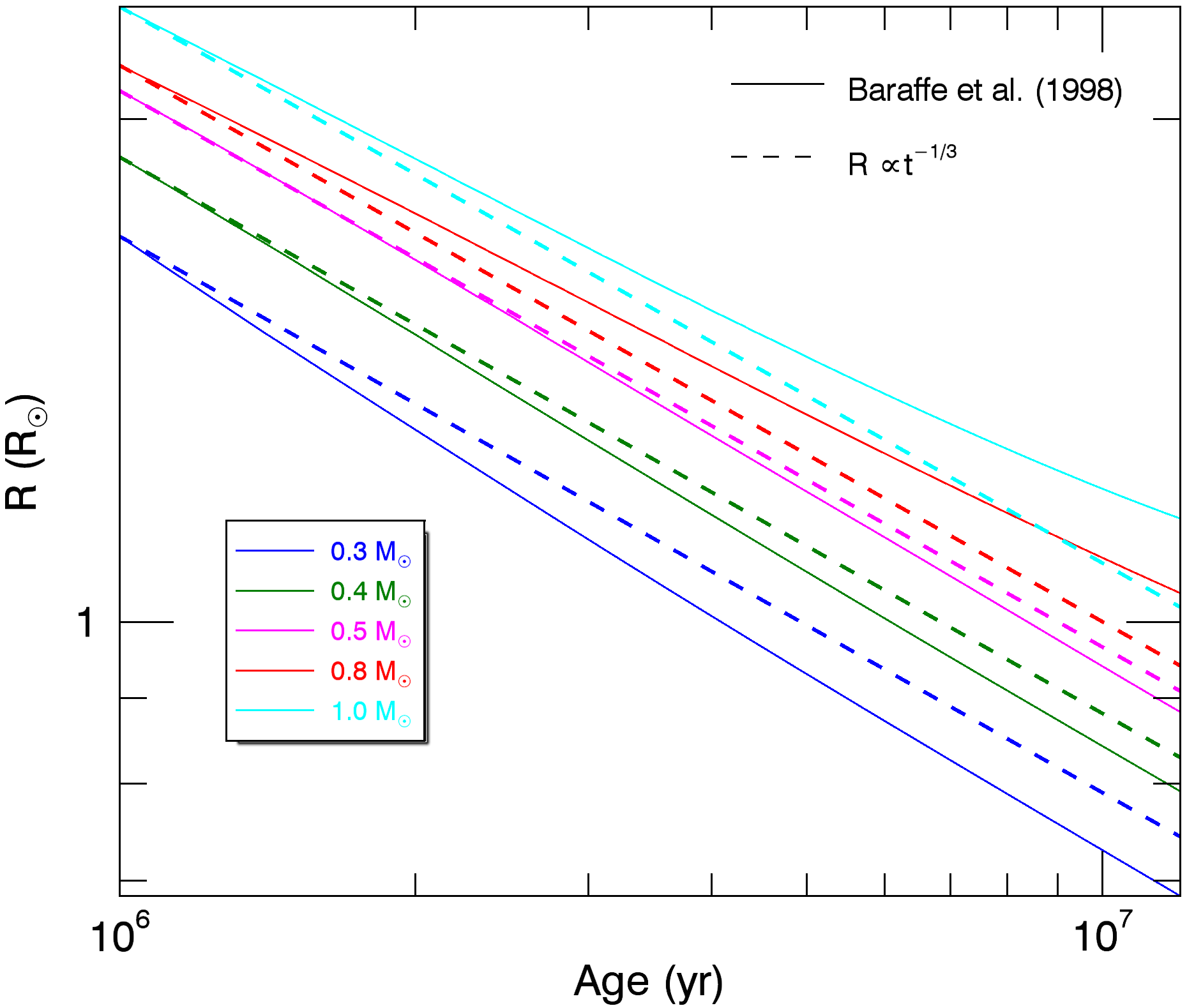}
\end{center}
\caption{Gyration radius squared (left) and stellar radius in solar
radius (right) as a function of time obtained from the stellar
evolution models of \citetads{1998A&A...337..403B} for the mass
bins considered in this work (colored solid lines). The colored
dashed lines are obtained assuming the contraction of a fully
convective polytropic pre-main sequence star $(R \propto t^{-1/3})$.
\label{rad_kgyr}}
\end{figure*}

\subsubsection{Specific angular momentum evolution}

\citetads{2014MNRAS.444.1157D} calculated the specific angular
momentum evolution of a sample of fully convective stars in
Taurus-Auriga and in the ONC. They used new estimates of stellar
radii and ages and classified their sample as Class II or Class III
stars based on {\it Spitzer} IRAC fluxes. They found that the
decreasing rate of the specific angular momentum during Class II
phase is given by j$_\mathrm{star} \propto$ t$^{-\beta_2}$, with
$\beta_2 = 2.0 - 2.5$, which is faster than expected if the angular
velocity is maintained constant. They interpreted this result as
indicating that the braking rate of accreting young stars is larger
than expected from disk-locking alone. They also observe the same
time dependency for Class III objects, which they interpreted as
initially accreting stars being sequentially released from their
disks over a timescale of about 10 Myr.

Using model M2 we calculated the specific angular momentum distribution
of our population. As stated before, in our simulations, while a
star has a disk, its angular velocity is kept constant. Then, for
disk stars, we expect that the specific angular momentum,
$j_\mathrm{star} (t) = k^2 R^2 \omega$, varies with time as
$j_\mathrm{star} \propto t^{2(\alpha + \beta)}$, if $k \propto
t^\alpha$ and $R \propto t^{\beta}$, assuming that the time dependency
of the gyration radius $k$ and the stellar radius $R$ can be expressed
as power laws. Both are obtained from the stellar evolution models
of \citetads{1998A&A...337..403B} and are shown in Figure \ref{rad_kgyr}
for each mass bin considered in this work. From 1 Myr to $\sim 3$
Myr, the gyration radius is approximately constant for all mass
bins.  At most, the gyration radius varies 1\% for the 0.3 M$_\odot$
mass bin. Beyond 3 Myr, the more massive stars start to develop a
radiative core and the gyration radius of the 1 M$_\odot$ decreases
by about 15\% at an age of 10 Myr, while the reduction is much less
in lower mass stars. For the asymptotic power-law form above, we
thus adopt $\alpha \simeq 0$, while in the simulations we use the
actual, time-dependent values of the gyration radius provided by
the stellar evolution models.

The variation of the stellar radius during the first few Myr mimics
that of a fully convective polytrope (n=1.5), i.e., $\beta = - 1/3$.
Fig.~\ref{rad_kgyr} show that for M = 0.3 - 0.5 M$_\odot$, the
stellar radius falls off slightly more rapidly with time ($\beta <
-1/3$), while the opposite is true for more massive stars. The exact
values of $\beta$, as derived from the stellar evolutionary models,
are shown in Table \ref{valbeta}. The mean value, however, is around
-0.33 and we will consider that $\beta = -0.33$ for the asymptotic
form above. Then, the expected time dependency of the specific
angular momentum in our simulations for disk-locked stars would be
$j \propto t^{-0.66}$. For diskless stars, the angular momentum is
held constant.

\begin{table}[t]
\caption{$\beta$ exponents from Figure \ref{rad_kgyr}  ($R_\star
\propto t^{\beta}$)}
\label{valbeta}
\centering
\begin{tabular}{cccccc}
\hline\hline
Mass (M$_\odot$) & 0.3 & 0.4 & 0.5 & 0.7 & 1.0 \\
\hline
$\beta$ & -0.36 & -0.35 & -0.35 & -0.29 & -0.28 \\
\hline
\end{tabular}
\end{table}

\begin{figure*}
\begin{center}
\includegraphics[width=16cm]{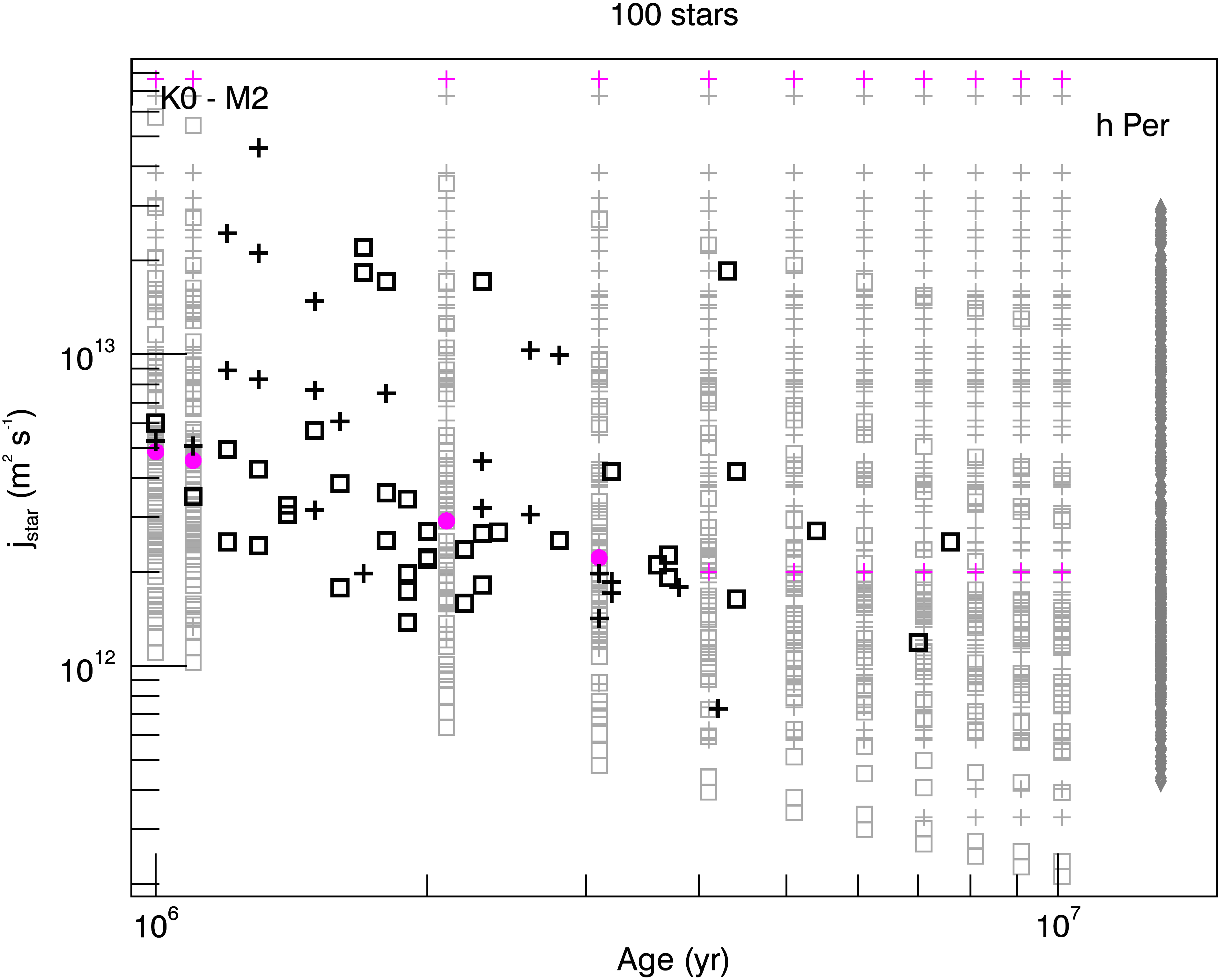}
\end{center}
\caption{The specific angular momentum evolution for 100 stars
randomly extracted from our simulations over the mass range
0.4-1.0~M$_\odot$ is compared to observed samples. Simulated data
are shown as light grey symbols, squares for disk stars and crosses
for diskless ones. Two stars chosen from the sample are highlighted
as magenta filled circles (disk) or crosses (diskless) to illustrate
the individual evolution depending on disk lifetime and initial
period. Observations: Taurus-Auriga and ONC (black symbols) samples
are from \citetads{2014MNRAS.444.1157D}, whose $j$ values were
recomputed taking the gyration radius from \citetads{1998A&A...337..403B}'s
models.  Squares represent Class II stars while crosses represent
Class III stars.  At 13 Myr, the specific angular momenta of
h Per members from \citetads{2013A&A...560A..13M} are plotted as
grey diamonds. \label{Jam}}
\end{figure*}

In Figure \ref{Jam}, the specific angular momentum, $j_\mathrm{star}$,
is plotted as a function of time for our simulated sample and
compared to observed samples of young stars.  At t = 1 Myr, $\sim$
75\% of the simulated sample consists of disk stars. The magenta
symbols show the specific angular momentum of two objects chosen
from our simulations. One is initially fast rotating and diskless,
and thus evolves conserving its initial angular momentum. The other
is initially a slower rotator and has a disk lifetime of about 3
Myr. During this time, its specific angular momentum decreases as
$t^{-0.66}$ as expected, while it is held constant once the star
is eventually released from its disk. When considered as a group,
we find that the median specific angular momentum of disk stars in
our simulation decreases with time as $j_{med} \propto t^{-0.65}$,
i.e., close to the expectation for constant angular velocity. For
diskless stars, we derive $j_{med} \propto t^{-0.53}$. Thus, while
individual diskless stars evolve at constant angular momentum, when
considered as a group their median angular momentum decreases in
time nearly as fast as for disk stars.  As already noted by
\citetads{2014MNRAS.444.1157D}, this is the result of the stars
being sequentially released from their disk over a wide range of
disk lifetimes, from about 1 to 10 Myr.

\begin{figure}
\begin{center}
\resizebox{\hsize}{!}{\includegraphics{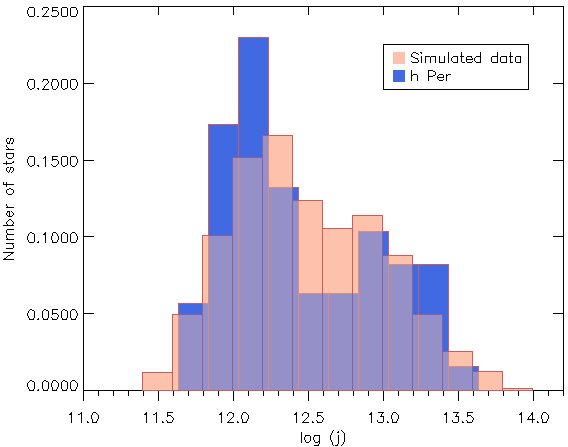}}
\end{center}
\caption{Distributions of the logarithm of the specific angular
momentum of our full sample at 12.1 Myr (red) obtained with model
M2 and of h Per members (blue). These last values were extracted
from \citetads{2013A&A...560A..13M}.
\label{comparhPer}}
\end{figure}

We compare our simulations to the $j_\mathrm{star}$ distribution
of the Taurus-Auriga and Orion samples from
\citetads{2014MNRAS.444.1157D}\footnote{Note that to allow for the
comparison, we had to recompute their values of $j$ using actual
gyration radii obtained from the \citeauthor{1998A&A...337..403B}'s
models.  \citetads{2014MNRAS.444.1157D} used for most of their
sample $k^2 = 2/3$ which applies to uniform density shells, while
a value of $k^2 = 0.205$ is appropriate for fully convective PMS
stars. We also neglected centrifugal effects in our simulations as
they don't impact significantly on the results.}. The Taurus-Auriga
and ONC $j$ estimates fall  well within our range of values. We
also plot the angular momentum values of h Per low mass members as
derived by \citetads{2013A&A...560A..13M} at an age of 13 Myr,
benchmarking the end of the accretion phase. A more detailed
comparison of the angular momentum distribution of h Per members
and that predicted by our simulations at 12.1 Myr is shown in Figure
\ref{comparhPer}.  While the distributions show differences,
especially in the location of the peaks, the overall range of
specific angular momentum is well accounted for. Indeed, this
suggests that the angular momentum distribution of low-mass stars
at the end of the PMS accretion phase builds up during the first
few Myr, as the result of disk locking acting over a wide range of
disk lifetimes, effectively widening the  initial $j$ distribution
(\citeads{1993A&A...272..176B}; \citeads{2004AJ....127.1029R}).

\section{Conclusions \label{conclusions}}

The Monte Carlo simulations presented here reproduce the main
observed rotational properties of young low mass stars during their
pre-main sequence evolution. We have shown that starting from now
well-documented initial distributions of periods, mass-accretion
rates and disk lifetimes, a model that assumes  disk locking for
accreting stars and angular momentum conservation for diskless stars
succeeds in reproducing both the evolution of rotational period
distributions in young clusters and the accretion-rotation connection
observed between IR excess and rotational period at young ages.  In
order to also reproduce the mass-rotation connection, one has to
further assume that disk locking is less efficient in very low mass
stars, below 0.3~M$_\odot$. Finally, the Monte Carlo simulations
naturally produce the distribution of angular momentum observed at
the end of the accretion phase, i.e., past 10 Myr. We show that
this distribution, which is the starting point for the subsequent
ZAMS and MS rotational evolution of low mass stars, builds up during
the PMS as the disk locking process acts over a wide range of disk
lifetimes, thus effectively widening the initial distribution of
periods during PMS evolution.

\begin{acknowledgements}

MJV is grateful to Florian Gallet and to Claire Davies for providing
theoretical and observational data to this work, to Adriano Hoth
Cerqueira and Andre Luis Batista Ribeiro for very useful discussions,
and to Florian Gallet and to Adriano Hoth Cerqueira for a careful
reading of the manuscript.  MJV would like also to thank the financial
support provided by CAPES (fellowship n. 2565-13-7) under the program
``Science without borders" and by the project PROCAD – CNPq/CAPES
number 552236/2011-0. JB acknowledges the support of ANR grant 2011
Blanc SIMI5-6 020 01 {\it Toupies: Towards understanding the spin
evolution of stars} (http://ipag.osug.fr/Anr\_Toupies/).

\end{acknowledgements}

\bibliographystyle{aa} 
\bibliography{ref3} 

\begin{thebibliography}{25}
\expandafter\ifx\csname natexlab\endcsname\relax\def\natexlab#1{#1}\fi

\bibitem[{{Affer} {et~al.}(2013){Affer}, {Micela}, {Favata}, {Flaccomio}, \&
  {Bouvier}}]{2013MNRAS.430.1433A}
{Affer}, L., {Micela}, G., {Favata}, F., {Flaccomio}, E., \& {Bouvier}, J.
  2013, \mnras, 430, 1433

\bibitem[{{Baraffe} {et~al.}(1998){Baraffe}, {Chabrier}, {Allard}, \&
  {Hauschildt}}]{1998A&A...337..403B}
{Baraffe}, I., {Chabrier}, G., {Allard}, F., \& {Hauschildt}, P.~H. 1998, \aap,
  337, 403

\bibitem[{{Bell} {et~al.}(2013){Bell}, {Naylor}, {Mayne}, {Jeffries}, \&
  {Littlefair}}]{2013MNRAS.434..806B}
{Bell}, C.~P.~M., {Naylor}, T., {Mayne}, N.~J., {Jeffries}, R.~D., \&
  {Littlefair}, S.~P. 2013, \mnras, 434, 806

\bibitem[{{Bouvier}(2013)}]{2013EAS....62..143B}
{Bouvier}, J. 2013, in EAS Publications Series, Vol.~62, EAS Publications
  Series, 143--168

\bibitem[{{Bouvier} {et~al.}(1993){Bouvier}, {Cabrit}, {Fernandez}, {Martin},
  \& {Matthews}}]{1993A&A...272..176B}
{Bouvier}, J., {Cabrit}, S., {Fernandez}, M., {Martin}, E.~L., \& {Matthews},
  J.~M. 1993, \aap, 272, 176

\bibitem[{{Bouvier} {et~al.}(2014){Bouvier}, {Matt}, {Mohanty}, {Scholz},
  {Stassun}, \& {Zanni}}]{2014prpl.conf..433B}
{Bouvier}, J., {Matt}, S.~P., {Mohanty}, S., {et~al.} 2014, Protostars and
  Planets VI, 433

\bibitem[{{Cieza} \& {Baliber}(2007)}]{2007ApJ...671..605C}
{Cieza}, L. \& {Baliber}, N. 2007, \apj, 671, 605

\bibitem[{{Dahm} \& {Hillenbrand}(2007)}]{2007AJ....133.2072D}
{Dahm}, S.~E. \& {Hillenbrand}, L.~A. 2007, \aj, 133, 2072

\bibitem[{{Davies} {et~al.}(2014){Davies}, {Gregory}, \&
  {Greaves}}]{2014MNRAS.444.1157D}
{Davies}, C.~L., {Gregory}, S.~G., \& {Greaves}, J.~S. 2014, \mnras, 444, 1157

\bibitem[{{Gallet} \& {Bouvier}(2013)}]{2013A&A...556A..36G}
{Gallet}, F. \& {Bouvier}, J. 2013, \aap, 556, A36

\bibitem[{{Hartmann} {et~al.}(1998){Hartmann}, {Calvet}, {Gullbring}, \&
  {D'Alessio}}]{1998ApJ...495..385H}
{Hartmann}, L., {Calvet}, N., {Gullbring}, E., \& {D'Alessio}, P. 1998, \apj,
  495, 385

\bibitem[{{Henderson} \& {Stassun}(2012)}]{2012ApJ...747...51H}
{Henderson}, C.~B. \& {Stassun}, K.~G. 2012, \apj, 747, 51

\bibitem[{{Hern{\'a}ndez} {et~al.}(2008){Hern{\'a}ndez}, {Hartmann}, {Calvet},
  {Jeffries}, {Gutermuth}, {Muzerolle}, \& {Stauffer}}]{2008ApJ...686.1195H}
{Hern{\'a}ndez}, J., {Hartmann}, L., {Calvet}, N., {et~al.} 2008, \apj, 686,
  1195

\bibitem[{{Hern{\'a}ndez} {et~al.}(2007){Hern{\'a}ndez}, {Hartmann}, {Megeath},
  {Gutermuth}, {Muzerolle}, {Calvet}, {Vivas}, {Brice{\~n}o}, {Allen},
  {Stauffer}, {Young}, \& {Fazio}}]{2007ApJ...662.1067H}
{Hern{\'a}ndez}, J., {Hartmann}, L., {Megeath}, T., {et~al.} 2007, \apj, 662,
  1067

\bibitem[{{Kroupa} {et~al.}(2013){Kroupa}, {Weidner}, {Pflamm-Altenburg},
  {Thies}, {Dabringhausen}, {Marks}, \& {Maschberger}}]{2013pss5.book..115K}
{Kroupa}, P., {Weidner}, C., {Pflamm-Altenburg}, J., {et~al.} 2013, {The
  Stellar and Sub-Stellar Initial Mass Function of Simple and Composite
  Populations}, ed. T.~D. {Oswalt} \& G.~{Gilmore}, 115

\bibitem[{{Lamm} {et~al.}(2005){Lamm}, {Mundt}, {Bailer-Jones}, \&
  {Herbst}}]{2005A&A...430.1005L}
{Lamm}, M.~H., {Mundt}, R., {Bailer-Jones}, C.~A.~L., \& {Herbst}, W. 2005,
  \aap, 430, 1005

\bibitem[{{Mamajek}(2009)}]{2009AIPC.1158....3M}
{Mamajek}, E.~E. 2009, in American Institute of Physics Conference Series, Vol.
  1158, American Institute of Physics Conference Series, ed. T.~{Usuda},
  M.~{Tamura}, \& M.~{Ishii}, 3--10

\bibitem[{{Manara} {et~al.}(2012){Manara}, {Robberto}, {Da Rio}, {Lodato},
  {Hillenbrand}, {Stassun}, \& {Soderblom}}]{2012ApJ...755..154M}
{Manara}, C.~F., {Robberto}, M., {Da Rio}, N., {et~al.} 2012, \apj, 755, 154

\bibitem[{{Moraux} {et~al.}(2013){Moraux}, {Artemenko}, {Bouvier}, {Irwin},
  {Ibrahimov}, {Magakian}, {Grankin}, {Nikogossian}, {Cardoso}, {Hodgkin},
  {Aigrain}, \& {Movsessian}}]{2013A&A...560A..13M}
{Moraux}, E., {Artemenko}, S., {Bouvier}, J., {et~al.} 2013, \aap, 560, A13

\bibitem[{{Rebull} {et~al.}(2006){Rebull}, {Stauffer}, {Megeath}, {Hora}, \&
  {Hartmann}}]{2006ApJ...646..297R}
{Rebull}, L.~M., {Stauffer}, J.~R., {Megeath}, S.~T., {Hora}, J.~L., \&
  {Hartmann}, L. 2006, \apj, 646, 297

\bibitem[{{Rebull} {et~al.}(2004){Rebull}, {Wolff}, \&
  {Strom}}]{2004AJ....127.1029R}
{Rebull}, L.~M., {Wolff}, S.~C., \& {Strom}, S.~E. 2004, \aj, 127, 1029

\bibitem[{{Ribas} {et~al.}(2014){Ribas}, {Mer{\'{\i}}n}, {Bouy}, \&
  {Maud}}]{2014A&A...561A..54R}
{Ribas}, {\'A}., {Mer{\'{\i}}n}, B., {Bouy}, H., \& {Maud}, L.~T. 2014, \aap,
  561, A54

\bibitem[{{Rigliaco} {et~al.}(2011){Rigliaco}, {Natta}, {Randich}, {Testi}, \&
  {Biazzo}}]{2011A&A...525A..47R}
{Rigliaco}, E., {Natta}, A., {Randich}, S., {Testi}, L., \& {Biazzo}, K. 2011,
  \aap, 525, A47

\bibitem[{{Spezzi} {et~al.}(2012){Spezzi}, {de Marchi}, {Panagia},
  {Sicilia-Aguilar}, \& {Ercolano}}]{2012MNRAS.421...78S}
{Spezzi}, L., {de Marchi}, G., {Panagia}, N., {Sicilia-Aguilar}, A., \&
  {Ercolano}, B. 2012, \mnras, 421, 78

\bibitem[{{Venuti} {et~al.}(2014){Venuti}, {Bouvier}, {Flaccomio}, {Alencar},
  {Irwin}, {Stauffer}, {Cody}, {Teixeira}, {Sousa}, {Micela}, {Cuillandre}, \&
  {Peres}}]{2014A&A...570A..82V}
{Venuti}, L., {Bouvier}, J., {Flaccomio}, E., {et~al.} 2014, \aap, 570, A82

\end{thebibliography}

\end{document}